\documentclass[12pt,preprint]{aastex}

\begin{document} 
\title{The Leo Elliptical NGC 3379:  A Metal-Poor Halo Emerges\footnote{
Based on observations made with the NASA/ESA Hubble Space Telescope, 
obtained at the Space Telescope Science Institute, which is 
operated by the Association of Universities for Research in Astronomy, Inc., 
under NASA contract NAS 5-26555. These observations are associated with program \#9811.
Support for this work was provided in part by NASA through grant 
number HST-GO-09811.01-A from the Space 
Telescope Science Institute, which is operated by the Association of 
Universities for Research in Astronomy, Inc., under NASA contract NAS 5-26555.}
}

\author{William E.~Harris} 
\affil{Department of Physics \& Astronomy, McMaster University, Hamilton L8S 4M1, Canada}
\email{harris@physics.mcmaster.ca} 

\author{Gretchen L.~H.~Harris}
\affil{Department of Physics \& Astronomy, University of Waterloo, Waterloo N2L 3G1, Canada}
\email{glharris@astro.uwaterloo.ca}

\author{Andrew C.~Layden}
\affil{Department of Physics and Astronomy, Bowling Green State University, 104 Overman Hall, Bowling Green, OH 43403}
\email{layden@baade.bgsu.edu} 

\author{Elizabeth M.~H.~Wehner}
\affil{Department of Physics \& Astronomy, McMaster University, Hamilton L8S 4M1, Canada}
\email{wehnere@physics.mcmaster.ca} 

\shorttitle{The Halo of NGC 3379}
\shortauthors{Harris et al.}

\begin{abstract} 
We have used the ACS camera on HST to obtain $(V,I)$ photometry for 
5300 red-giant stars in the halo of the dominant Leo-group member NGC 3379, 
a galaxy usually regarded as a classic normal giant elliptical.  
We use this sample of stars
to derive the metallicity distribution function (MDF) for its outer-halo
field stars at a location centered 33 kpc from the galaxy center.  
In at least two ways the MDF is distinctly unlike all the other E galaxies for which
we have similar data (including the Local Group dwarf ellipticals,
the intermediate-luminosity NGC 3377, and the giant NGC 5128). First, 
the MDF for the NGC 3379 outer halo is extremely broad and flat, with many stars 
at every interval in [m/H] and only a gradual rise towards higher
metallicity.  Second, we see a metallicity gradient across our
ACS field such that in its outermost region the blue, low-metallicity
stars ([m/H] $< -0.7$) are beginning to dominate and the higher-metallicity
stars are rapidly diminishing.  In order to successfully match this
extremely broad MDF, we find that a distinct two-stage 
chemical evolution model is necessary.
Our target field is centered at a projected distance about equal
to $12 R_e$, twice as far out in units of effective radius as in any
of the other galaxies that we have surveyed. 
If NGC 3379 is indeed 
representative of large E/S0 galaxies, we predict that 
such galaxies in general will reveal diffuse low-metallicity subpopulations, but
that photometry at radii $r \simeq 10 - 15 R_e$ will be necessary
to get beyond the edge of the dominant metal-rich
component and to see the faint low-metallicity component clearly.
Finally, we discuss possible connections of these outer-halo stars with
the metallicity distributions that are beginning to be observed in
the intracluster medium of nearby rich galaxy clusters, which also
show flat MDFs.  
These outermost-halo observations are challenging, but ones which 
may give an unusually direct
window into the earliest star-forming stages of these galaxies.

\end{abstract}


\section{Introduction} 
\label{intro}

The imaging tools provided by the Hubble Space Telescope have revolutionized
our ability to study the stellar populations in nearby galaxies, at depths
and resolutions that have been impossible with ground-based instruments.
Particularly for the oldest stellar populations -- the halo stars, bulge
stars, and globular clusters -- these deep and precise HST-based photometric
studies have opened a major 
route to understanding the early chemical evolution of their host
galaxies that was previously available only for the Local Group members.

The giant elliptical galaxies attract special interest because
they may result from the widest possible range of formation histories,
all the way from hierarchical merging at very early times, to recent major mergers,
to later growth by satellite accretion.  
When considered along with the dwarf ellipticals within the Local Group (NGC 147, 185, 205, M32,
and the many dwarf spheroidals),
we can, at least in principle, piece together the evolutionary histories of E 
galaxies over their full mass range.

The nearest giant E/S0 galaxy is NGC 5128, the dominant member of the
Centaurus group at $d=3.8$ Mpc.  In a series of previous papers \citep{h99,h00,h02,rej05}, 
we have discussed photometric analyses
of its halo and bulge stars covering field
locations at projected distances ranging from 8 to 40 kpc.  In all
four of these studies the red-giant branch (RGB) stars are clearly resolved, and in the
deepest one \citep{rej05}, even the old horizontal-branch population is reached.  
Intriguingly, throughout every part of its halo that we have surveyed so far,
the {\sl metallicity distribution function} (MDF) of the giant stars
is clearly metal-rich (with a mode near [Fe/H] $\simeq -0.4$) and
with extremely small proportions of classically metal-poor stars in the range
[Fe/H] $< -1$.  However, uncertainties continually surround any attempt to
generalize NGC 5128 to all large ellipticals because of its evident
history of satellite accretion from within the Centaurus group
\citep[e.g.][]{is98,peng02,kaw06,malin83,tubbs80,quil93}.  The possibility
of a larger merger  has been modelled by \citet{bekki06}.
In addition, models show that both
a major-merger and a hierarchical-merging approach are capable of creating
an MDF with the same basic characteristic of a predominantly
metal-rich halo \citep{bek01,bekki03,bea03}, although the mechanisms in
each case are different:  in a major merger, the halo ends up being
populated by the metal-rich stars in the disks of the colliding
progenitors; while in hierarchical merging, the metal-rich stars 
accumulate in the long series of small and large starbursts that
construct the galaxy as a whole.  Even though many large ellipticals
share the same kinds of features with NGC 5128, it is necessary to extend
these studies to other targets to gain the complete picture that we need.

The next nearest readily accessible 
E galaxies are in the Leo group at $d \sim 10$ Mpc, including
the intermediate-luminosity NGC 3377 and the giant NGC 3379.  
The E1 giant NGC 3379 (Messier 105)
is an especially attractive target because it is, quite literally,
a textbook giant elliptical \citep[e.g.][]{carroll07}.  
It is a keystone \citep{devauc79}
in establishing the classic de Vaucouleurs photometric profile;
\citet{ss99} refer to it engagingly as ``virtually a walking advertisement
for the $r^{1/4}$ law''.  With a luminosity $M_V^T = -20.85$,
a smooth and nearly round profile shape, no strong photometric peculiarities,
and a nearby location in a high-latitude field, it has for decades been
regarded as a baseline ``normal'' elliptical.\footnote{A more generally
flexible model for matching a wider range of E galaxy profiles is the
generalized Sersic model \citep[e.g.][]{ferr06} or core-Sersic model
with index n depending on luminosity; the traditional de Vaucouleurs profile
is a special case of this family.}  A detailed study of its
halo stars is within reach of the HST ACS camera, and thus holds out considerable
promise for giving us new insight into the stellar populations of
classic giant ellipticals.  Table 1 summarizes its basic parameters.

NGC 3379 fits well into the E-galaxy sequence in other ways.  Its central
black hole mass of $M_{BH} \simeq 1.4 \times 10^8 M_{\odot}$
\citep{sha06} closely follows the normal $M_{BH} - \sigma$ relation.
Its velocity field and dynamical structure are well behaved out to
$R_{gc} \sim 90'' \simeq 2 R_e$ \citep{sha06}
and large-scale surface photometry shows no shells or other remnants
of mergers or accretions \citep{sch92} and very little gas 
\citep[see][for a review]{ss99}.
The possibility has been discussed that it may actually be an S0 or modestly 
triaxial configuration seen nearly face-on, since the right combinations
of disk, bulge, and inclination angle could mimic a global $r^{1/4}$
profile \citep{ss99,cap91,sha06}.  The velocity field within $R < R_e$
may also be more consistent with an S0 structure \citep{ss99}.
In the discussion below, other evidence will be mentioned that may 
also be consistent with an S0 classification.

Previous studies of the resolved old red giant stellar populations in
NGC 3379 have been published by \citet{sakai97} and \citet{gregg04}.
\citet{sakai97} used HST WFPC2 exposures in a field $6'$ west of galaxy
center in a single filter (the $F814W$ ``wide $I$'' band) in order to 
calibrate the distance from the tip of the red-giant branch.
\citet{gregg04} used HST {\sl NICMOS} $J$ and $H$ exposures in three fields,
one of them located within the Sakai WFPC2 field and the other two
further in, at $R_{gc} = 3'$ and $4\farcm5$ from galaxy center.  
They found that the mean metallicity for the RGB stars in these 
fields was near Solar abundance, but since metal-poor stars are significantly
fainter than metal-rich ones in the near infrared, they left open the
possibility that the mean [m/H] might be overestimated.  At fainter levels
in their color-magnitude diagrams, low-metallicity
stars with [m/H] $< -1$ may begin to appear, but the
larger photometric measurement scatter there leaves uncertainties about
the relative numbers versus metallicity.  At the opposite end
of the abundance scale,
stars of metallicity [m/H] $= +0.4$ and even higher could also be present
according to the range of colors they observe in the CMD.
Direct comparison of similar {\sl NICMOS} photometry for an inner-halo
field in NGC 5128 by \citet{mar00} indicates that the {\sl mean} 
stellar metallicities are both near Solar in these two galaxies
but that the internal metallicity spread in NGC 3379 is higher.
We emphasize, however, that both of these studies have targeted the
inner halos of these gE galaxies and thus must be sampling 
predominantly the bulge (or disk) population that is expected to be metal-rich.

In this paper, we present new color-magnitude photometry for the halo stars in
NGC 3379.  The observations and analysis techniques are the same as in
our companion study of the other Leo elliptical, NGC 3377 \citep{har07}.
Although we expected to find that the NGC 3379 halo would be metal-rich
following the pattern established by the other ellipticals already
studied, the results have turned out differently.

\section{Observations and Data Reduction}
\label{observations}

\begin{figure} 
\figurenum{1} 
\plotone{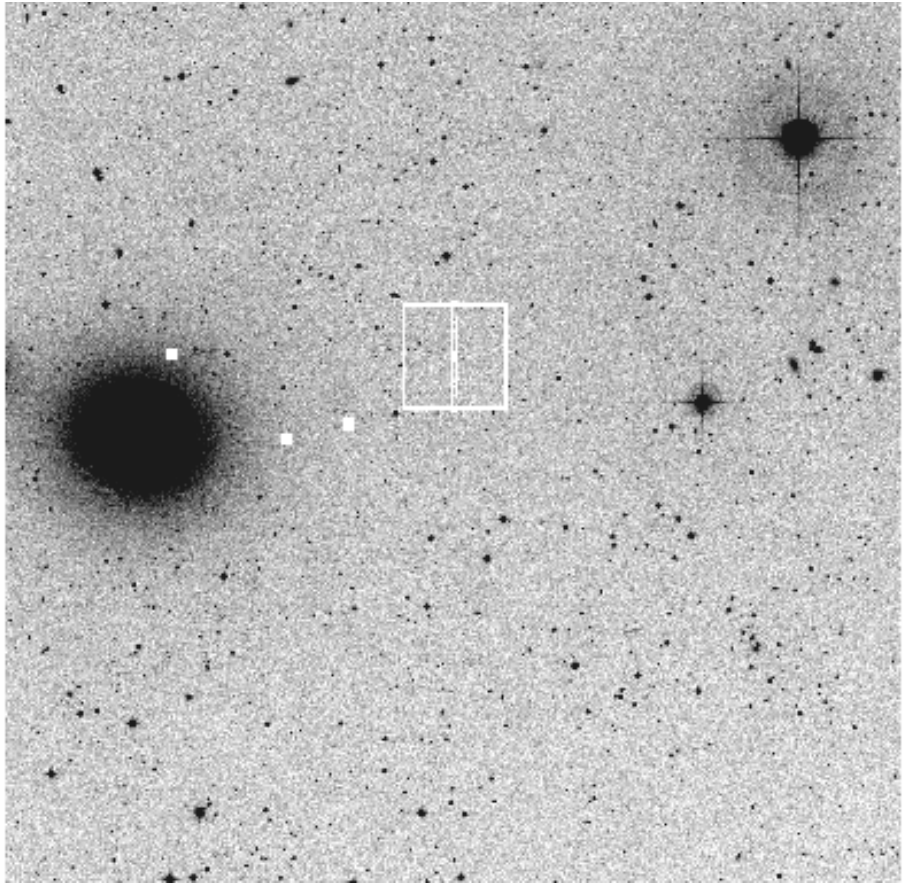} 
\caption{A wide-field image of the area containing our 
target field, extracted from the Digital Sky Survey.
The bright galaxy near the left edge is NGC 3379 and
the entire field is $30'$ across; North is at top and East at left.
Our ACS/WFC field is marked by the large box at the center of the frame;
the vertical bar through the middle shows the orientation of the gap
between the two WFC CCD detectors.  The three solid dots between
NGC 3379 and the ACS field show the locations of the three NICMOS fields
observed by Gregg et al.~(2004).  The WFPC2 field imaged in $I$ by
Sakai et al. (1997) is centered on the outermost of the three
NICMOS fields.
}
\label{widefield}
\end{figure}

\begin{figure} 
\figurenum{2} 
\plotone{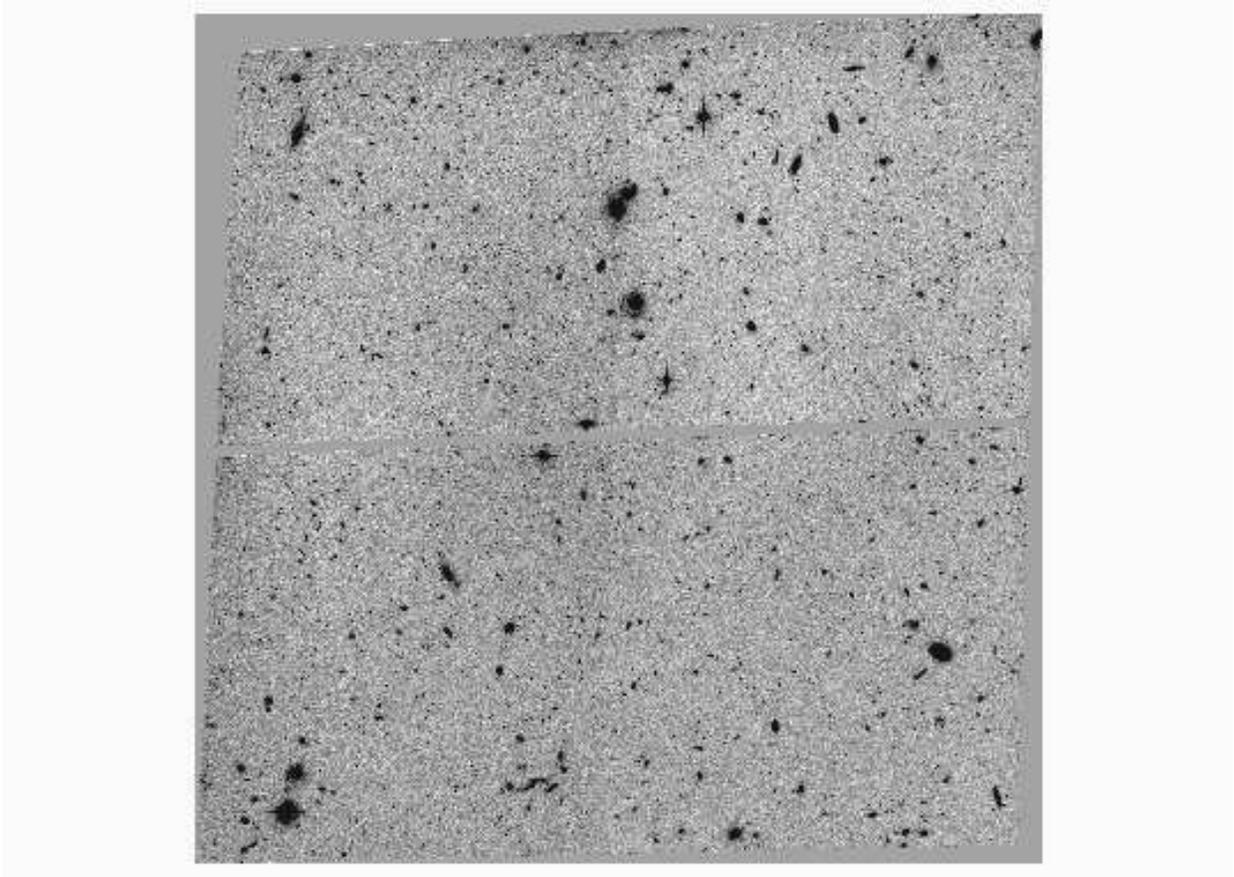} 
\caption{A reproduction of our ACS Wide Field Channel field.  
The orientation of this field is rotated nearly at right angles
relative to the previous wide-field image;
here, North is at the right and East at top.  The $y-$axis of
the ACS is directed $96^o$ E of N.
Notice the large number of background galaxies, which are the dominant
source of field contamination (see text).}
\label{acsfield}
\end{figure}

Our imaging data for both NGC 3377 and 3379 were obtained in HST program 9811. 
We used the Advanced Camera for Surveys in its Wide Field Channel,
which has an image scale $0\farcs05$ per pixel. 
Our NGC 3379 target field was
centered at $\alpha = 10^h 47^m 06\fm5, \delta = +12\arcdeg 37\arcmin 46\farcs9$
(J2000).  This field is $630''$ west and $173''$ north of the center
of NGC 3379, equivalent to
$R_{gc} = 10\farcm9 \simeq$33 kpc projected radius at our adopted distance of
10.2 Mpc (see below for the distance calibration).    
The galaxy light profile has an effective radius $R_e = 0\farcm93 \simeq 2.8$ kpc, 
putting our target field at $R_{gc} \simeq 11.7 R_e$.
We deliberately selected a location about twice as
far out as the WFPC2 field location used by \citet{sakai97} to avoid any
concerns about image crowding, as well as to ensure that we would be gathering
a sample of stars that could be viewed as a genuine ``halo'' population
different from the inner fields in these previous studies.
The field placement is shown in Figures \ref{widefield} and \ref{acsfield}.

The comprehensive surface-photometry studies of \citet{devauc79} and 
\citet{cap90} found that
NGC 3379 has isophotal contours with a mean ellipticity $\langle b/a\rangle = 0.88$ and
a major axis orientation $70^o$ E of N (or $110^o$ W of N).  Our target
field, relative to the galaxy center, lies in a direction $75^o$ W of N and
thus is at an angle of $35^o$ off the major axis.  
On the opposite (east) side of NGC 3379 is the disk galaxy NGC 3384, at
$\alpha = 10^h 48^m 16\fm9, \delta = +12\arcdeg 37\arcmin 46\arcsec$.
Our ACS field is thus nearly due west of NGC 3384 and $1030''$ away.
Since NGC 3384 is nearly twice as far away as NGC 3379, and only about
half as luminous, we expect the stellar population visible on our
target field to be completely dominated by the giant elliptical.

We used the ACS/WFC ``wide V'' ($F606W$) and ``wide I'' ($F814W$) filters,
the same ones as in our previous studies of NGC 5128 \citep{h99,h00,h02,rej05}
and NGC 3377 \citep{har07}.   As we discuss
in those papers, the
$(V-I)$ color index is an effective metallicity indicator for 
old red giant stars over the full
metallicity range from [Fe/H] $\sim -2$ up to Solar abundance, 
and particularly for [Fe/H] $\gtrsim -1.5$.  
Over this full metallicity range, the corresponding spread of
$(V-I)$ colors is more than 2 magnitudes at the
top of the giant branch (see the color-magnitude
diagrams plotted below), a range more than twice as large as
the $\sim 0.8$-mag color spread in near-infrared indices such as $(J-H)$
\citep{gregg04}.  The second major advantage that optical color
indices have over infrared ones is the very much bigger detector
area, so that that statistically larger sample sizes can be
accumulated much faster.

In the $F606W$ filter our
total exposure time was 38500 sec split over 15 exposures; for
$F814W$, the total was 22260 sec over 9 exposures. In the original
series of exposures, half the $F606W$ images were ruined by camera
shutter and readout failure; fortunately, these were repeated successfully
a year later.  Recovering the full series of images proved to be crucially
important to our main goal of interpreting the metallicity distribution 
function of the halo stars, because our detection of the reddest (thus
most metal-rich) RGB stars is set by the photometric limits in the $V$ filter.
The final exposures 
were identical with the totals for our NGC 3377 data, although
(as will be seen below) the NGC 3379 data reach slightly deeper 
because of decreased effects of crowding.
The individual exposures in the series were dithered over several step sizes up to   
20 pixels, allowing elimination of most cosmic rays, bad pixels, and other artifacts
on the detector.  To prepare the images for photometry, we extracted the drizzled
individual images from the HST Data Archive, registered them to within 0.05 pixels,
and median-combined them.  This procedure gave us a single very deep 
exposure in each filter.  In Figure 2, we show the combined $I-$band image. 

Our procedures for object detection and photometry were identical
with those for NGC 3377 \citep{har07} and are more fully described there.
In summary, we used the standalone version of {\sl DAOPHOT} codes in
its fourth-generation {\sl daophot 4} version, with the
normal sequence of {\sl find/phot/allstar}.
The primary difference during the {\sl daophot} stage compared with our
companion NGC 3377 study
was that the number density of stars 
was about one full order of magnitude {\sl lower} than on our NGC 3377 field,
so that we had to select candidate bright, isolated stars to define the
point spread function by visual inspection to weed out the many similarly bright
but nonstellar background galaxies.  In the end, the PSF was defined from an average of 15 to
20 stars on each frame.  The FWHM of the point spread function is
2.3 px or $0\farcs115$. 

The detected objects on each of the two master images
were matched up to isolate those measured in both colors.  At this stage,
any objects with {\sl allstar} goodness-of-fit parameters
$\chi_V > 1.5$ or $\chi_I > 1.2$ were rejected, leaving a total
of 5642 matches.  Not all of these, however, are stars.  The biggest single issue we had
to deal with in this dataset was not crowding or faintness; in absolute terms
this outer-halo field is completely uncrowded, and the brightest red giants
in the galaxy are quite well resolved.  Instead, the main problem was {\sl field contamination}
by faint background galaxies, which appeared in larger numbers on this particular
region of sky and made up a relatively much higher proportion of
the total population of objects on the frame than in our
NGC 3377 field.  To define the cleanest possible subset of data, first we
masked out circles around more than a hundred of the biggest field galaxies 
in the field, which are ``detected'' by {\sl daophot/find} as multiple
closely spaced clusters of PSFs, all of which are invalid.  This step eliminated
10 percent of the total field area.  Next, we ran the SExtractor object
detection and classification code \citep{bertin} and rejected any objects with
``stellarity'' index less than 0.3 (although many of these had in fact already
been rejected in the {\sl daophot/find} and {\sl allstar} steps).  Finally, we used extremely
careful visual inspection of all the remaining objects brighter than
$I \simeq 27.5$ (which is about 1.5 mag below the red-giant-branch tip;
see below) to pick out any other definitely nonstellar or closely crowded
objects.  This series of steps left us with a final total of 5323 stars.
For comparison, we obtained a total of 57039 stars in NGC 3377 
over the same ACS field area, where the pointing was relatively
closer to the galaxy center.

Calibration of the photometry followed exactly the same steps
described in \citet{har07}.  The adopted transformations
between the $F606W$ and $F814W$ magnitudes on the 
natural ACS VEGAMAG filter system, and the standard $V$ and $I$, are
repeated here for convenience:
\begin{eqnarray}
F606W \, = \, V - 0.265 (V-I) + 0.025 (V-I)^2 \\
F814W \, = \, I + 0.028 (V-I) - 0.008 (V-I)^2
\end{eqnarray}

We next carried out conventional artificial-star tests to determine
the internal photometric uncertainties and the detection
completeness.  Scaled PSFs were added to the master images 
1000 at a time over a wide range of magnitudes,
independently on the $F606W$ and $F814W$ images,  and the
images were then remeasured in exactly the same way
as the original frames.  A total of 10,000 fake stars were used
in $V$ and 8000 in $I$.  For relatively uncrowded frames such
as these, the fraction $f$ of stars recovered,
as a function of instrumental magnitude, is well described by 
a Pritchet interpolation curve   
\citep{fl95}, 
\begin{equation}
f=\frac{1}{2}[1-\frac{\alpha(m-m_0)}{\sqrt{1+\alpha^{2}(m-m_0)^{2}}}]
\end{equation}
which has two free parameters: the limiting magnitude $m_0$ where $f=0.5$, and the slope
$\alpha$ giving the steepness of the $f-$dropoff through the $m_0$ point.
For our data we find $m_0(F606W) = 29.20$ and $m_0(F814W) = 28.10$,  along with
$\alpha(F606W)=2.8$, $\alpha(F814W)=3.0$.
These limits are both $0.25 - 0.3$ mag deeper than in our NGC 3377 field.

The artificial-star tests can also be used to estimate the internal random
uncertainties of the photometry, as well as any systematic bias in the measured
magnitudes as functions of magnitude.  No biasses larger than 0.03 mag in
either filter were found for stars brighter than the completeness limit $m_0$,
and the resulting biasses in the color indices ($V-I$ or $F606W-F814W$)
are completely negligible.
The mean random uncertainties are represented accurately by gradually
increasing exponential interpolation curves,
\begin{eqnarray}
\sigma(F606W) \, = \, 0.01 + 0.03\, {\rm exp}((F606W - 27.0)/1.09) \\
\sigma(F814W) \, = \, 0.01 + 0.03\, {\rm exp}((F814W - 26.0)/1.15) \, .
\end{eqnarray}
Over our primary magnitude range of interest ($I \lesssim 27$) the measurement
uncertainties are less than $\pm 0.1$ mag, much less than the intrinsic
spread in colors for the bright RGB stars we are studying.  Although these
interpolation equations indicated that the
{\sl internal} precision of the photometry is only $\simeq 0.01$ mag at the
bright end, the true (external) uncertainty could be $\pm 0.02 - 0.3$ mag
because of other factors such as systematic variations of the PSF across the
field and accuracy of flat-fielding.  However, photometric uncertainties at
any such level are trivial compared with the $> 1$-mag range in $(V-I)$ colors
that we use for metallicity determination.

An important feature of the completeness limits is that the limiting
curve for $V$ cuts off
our ability to see any extremely red stars that might actually be present;
these would fall at the most metal-rich end of our metallicity distribution
function.  Considerably deeper exposures in $V$ will be needed to explore the true
``red limit'' of the giant stars in this galaxy.  Within the limits imposed by
the photometry, we explicitly take into account the completeness fraction $f$ in
our derivation (below) of the metallicity distribution.  
Near-infrared photometry would be sensitive to this high-end range of metallicities
\citep{gregg04}, but as noted earlier, the accompanying huge penalty is the
smaller area of the infrared detectors and loss of statistical weight in
sample size.

\section{The Color-Magnitude Diagram}
\label{distance}

\begin{figure} 
\figurenum{3} 
\plotone{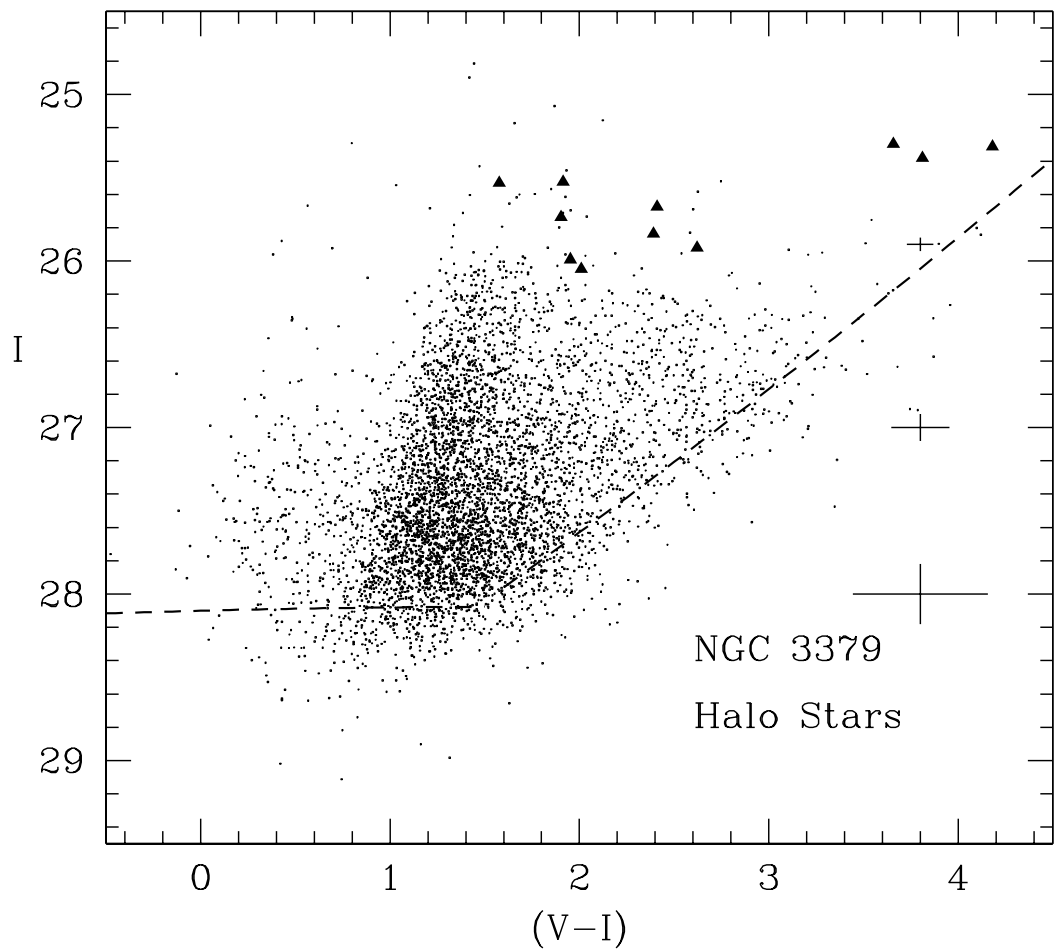} 
\caption{The color-magnitude array in $I, (V-I)$ for our measured sample
of 5323 stars in the halo of NGC 3379.  The dashed lines show the
50\% detection completeness limits in $I$ (horizontal line at bottom)
and $V$ (line sloping upward to the right).  The errorbars shown along
the right side are for an average color $(V-I) = 1.5$.
Eleven candidate Long Period Variables (LPVs) detected from the
two separate epochs of the $V$-band exposures are marked with 
large triangles (see Section 3 of the text).
}
\label{rawcmd}
\end{figure}

The color-magnitude diagram (CMD) of our final sample of 5323 stars is
shown in Figure \ref{rawcmd}.  The presence of a substantial
{\sl blue} RGB population in the color range $1 \lesssim (V-I) \lesssim 1.6$
is immediately evident, along with many redder RGB stars that continue
redward till the $F606W$ completeness cutoff line (dashed line at right).
The first impression  is therefore 
that the halo has primarily a {\sl blue, metal-poor}
giant branch like those in dwarf ellipticals \citep[e.g.][]{han97,butler05}.
However, this 
initial reaction about the relative numbers of stars at various
metallicities is deceptive because of the strongly nonlinear
dependence of color on metallicity, as will be discussed 
in Section \ref{metallicity} below.

The region to the blue side of the RGB (those with $(V-I) < 0.9$ 
and $I > 27$) also has some objects scattered across it, and
is more well populated than in our NGC 3377 data.
If real, it might suggest the presence of a young population.  However, the
great majority of these blue objects appear,
once again, to be due simply to field contamination.  A plot of the 
$xy$ positions of these very blue objects shows that they
are rather uniformly scattered over the field, as would be expected
for a background or foreground population (see the discussion in 
Section \ref{gradient} below for an analysis of spatial
gradients in the RGB stars), although a shallow gradient would
be hard to detect with this small number of stars.  
A more incisive test is from direct examination 
of our images, which suggests that
most of the very blue objects are noticeably nonstellar, or crowded, or both.  However,
for objects this faint it is extremely difficult for objective 
routines such as {\sl daophot} or SExtractor, or even careful eye inspection,
to classify and separate them cleanly from stars.  In this range as well,
the photometric measurement errors become large enough to contribute
a noticeable spread in the RGB color distribution, as is evident
in the CMD.  

Pursuing the tests for a younger population a bit further, we have
experimented with the placement of isochrones of various ages on
the CMD.  If these blue objects are young, moderately metal-rich stars,
they would have to be near $10^8$ yr and older to pass through the relevant
region of the CMD.  A population of such stars should also produce an
AGB branch extending up above the RGB tip, which does not appear
\citep[cf.][for examples of this type of feature]{wil07}.
In addition, star formation within the past $10^8$ yr could also be
expected to leave residual gas, for which little evidence exists.
In sum, it seems appropriate to conclude that most of these objects
are simply field contamination. 

The classification steps that
we carried out (see above) are much more definitive for $I < 27$,
which fortunately is the range that we rely on for our results (the MDF and the
distance measurement).  In this key upper RGB range, we are working with a 
sample of well defined, isolated stars.  Over the color range $(V-I) \gtrsim 1.0$,
the stars can easily be interpreted as the halo RGB population over a range
of metallicities.

The last region of the CMD calling for particular comment is the
section brighter than the RGB tip (TRGB).  We have already noted that
no obvious younger AGB-like branch is present in this supra-TRGB
area, and of the
$\simeq$40 stars lying clearly above the TRGB, half of them
can be understood simply as foreground field-star contamination
(we expect $\simeq 20$ from Galactic starcount models; 
e.g. Bahcall \& Soneira 1981). 
The normal, old RGB can also produce some brighter objects, such
as LPVs in
temporary luminous stages \citep{ren98,har07}, and accidental blends
of RGB stars that are measured as brighter singles.  
LPVs and similar objects are present in proportion to the total
RGB population, while the number of accidental blends due to crowding
goes up as $N_{\star}^2$.  
For our similar studies of NGC 3377 \citep{har07} and the bulge region of
NGC 5128 \citep{h02}, the numbers of stars above the RGB tip were 
significantly larger, driven by the much higher density of stars.
For this very uncrowded NGC 3379 field, we expect $\simeq 10 - 20$ LPVs
but essentially no accidental blends.  In summary, the total of all expected
sources of supra-RGB sources matches the number we see to within
statistical uncertainty.

As a final approximate but
more direct test for the presence of LPVs, we used the fact that
the $F606W$ images were taken in two groups at widely separated epochs
to search directly for bright variables.
We mdeian-combined the first six $F606W$
exposures taken in 2004 to make a single,
cleaned image, and similarly combined the four additional $F606W$ exposures taken
almost exactly one year later to make a second cleaned image.  We then ran
the normal sequence of $\sl{find/phot/allstar}$ 
on these two images as described in Section \ref{observations} and
merged the photometry files using $\sl{daoamaster}$.  Only stars that
were retained in the culling steps described in Section \ref{observations} were
considered.  Stars showing a magnitude difference $\Delta V$ between the two
epochs (including stars found in one but not the other epoch) greater
than seven times their measurement uncertainties were considered 
candidate variables.  We
inspected these stars on the two images to verify that the magnitude
variations were not affected by incomplete filtering of cosmic rays
or other image artifacts.
This procedure gave us a list of eight LPV candidates.  We also visually
inspected all stars with $I < 26$ and $V-I > 3$ mag, since these red,
metal-rich stars are also good candidates for LPVs.  The three
brightest such stars appear distinctly in one image but not the other,
so are likely to be LPVs as well.  While several of the total of 11 LPV
candidates have magnitude differences $\Delta V > 1.0$ mag, others are
closer to the limit of detection.  Given that we have sampled the data
at only two distinct epochs as well, we therefore expect there are other LPVs in
the field that we were unable to detect. The 11 LPV candidates are marked
in the CMD of  Fig.~\ref{rawcmd}.

In summary, the number of candidate LPVs that we have identified is
consistent with our rough estimates above that were based only on the population
statistics of the RGB.  These 11 LPVs,
together with the $\simeq 20$ field stars, already account for most
or all of the subra-RGB sources seen in Fig.~\ref{rawcmd}, to within
statistical uncertainty.

\section{Distance Calibration}
\label{trgb}

Most of the large galaxies in the Leo I group are disk systems, while
NGC 3379 is the largest elliptical.  In \citet{har07}, we summarize
the previous measurements of distance
to individual Leo members from a variety of
well established distance indicators including Cepheids, planetary nebula
luminosity function (PNLF), surface brightness fluctuation (SBF), and
the tip of the old red-giant branch (TRGB).  The overall average for
5 large galaxies including both NGC 3377 and NGC 3379 \citep[see][]{har07}
is $\mu = (m-M)_0 = 30.1 \pm 0.05$, or $d = 10.4$ Mpc.
The galaxy-to-galaxy dispersion of these measurements
($\sigma_{\mu} = 0.17$ mag) is comparable with the internal uncertainties
of each method.

For NGC 3379 specifically, the TRGB method as applied
through WFPC2 photometry in the optical $I$ band \citep{sakai97} 
gave $\mu = 30.30 \pm 0.27$, while the same method from HST/NICMOS in
the near infrared \citep{gregg04} gave $\mu = 30.17 \pm 0.12$.  
The PNLF method \citep{ciar89} yielded
$\mu = 29.96 \pm 0.16$, and the SBF method \citep{tonry01} $\mu=30.12$.

Our new ACS photometry penetrates well into the the old-halo red giant branch
with a cleanly defined sample,
and provides a new opportunity to use the TRGB distance indicator more precisely
than before.  The brightest detectable RGB stars, by hypothesis, define
the ``tip magnitude'' or TRGB, which represents
the luminosity of the helium flash at the end of the stars'
first ascent along the giant branch.  Empirically, we plot 
the luminosity function of the RGB stars in the $I$ band and use the sharp rise in the LF
to define the onset of the RGB.  The method is outlined by 
\citet{sakai96,sakai97} and \citet{h99} and these papers can be seen for
further discussion of the technique.
For stars more metal-poor than
[Fe/H] $\simeq -0.7$ (which include the majority of the ones we measure here;
see next section), the $I$ band has the strong advantage that the differential
bolometric correction across the top of the RGB is almost cancelled by
the opposite dependence of $M_{bol}(tip)$ on metallicity, leaving 
only a gradual decrease of $M_I(tip)$ with increasing color.

\begin{figure} 
\figurenum{4} 
\plotone{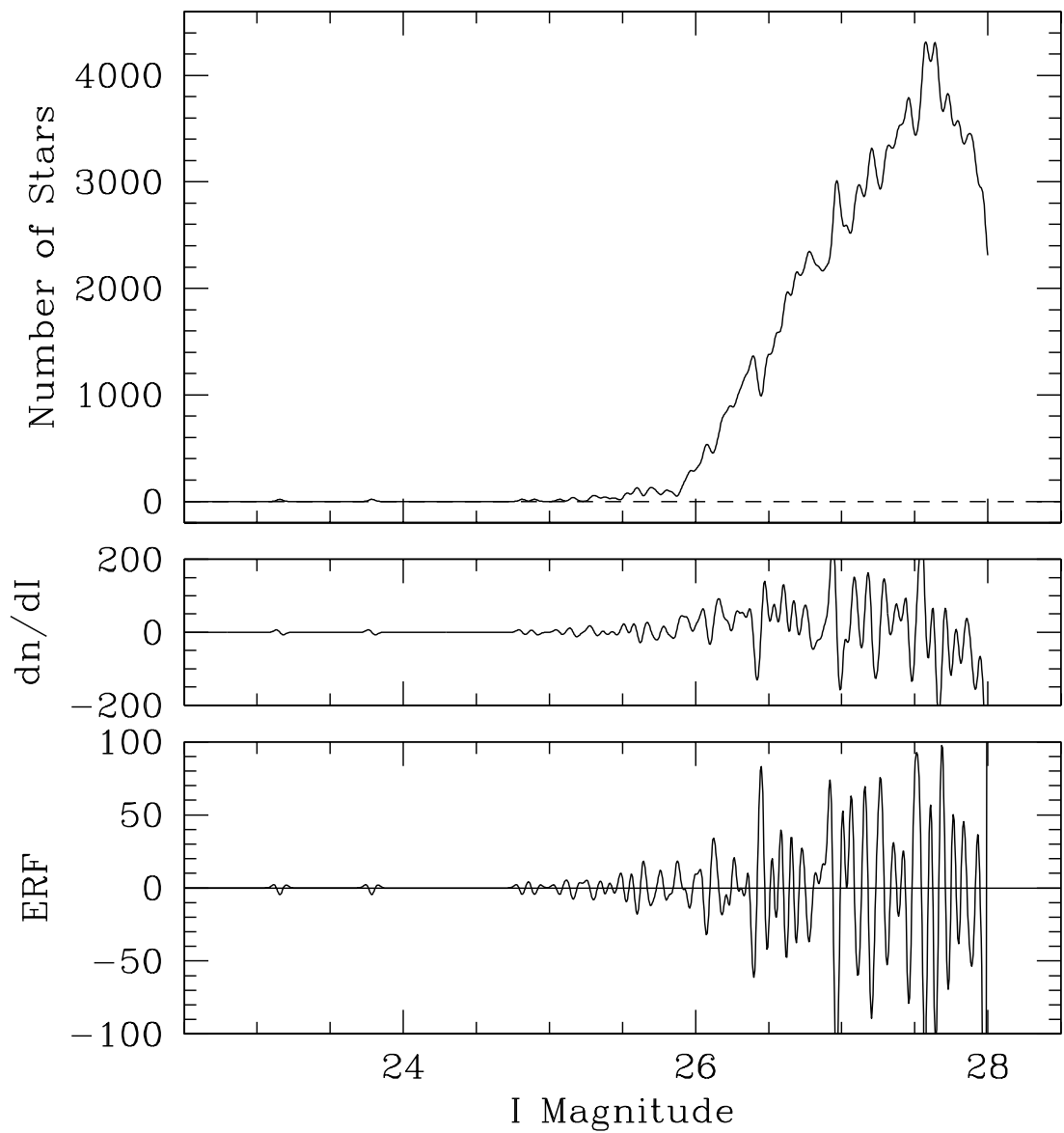} 
\caption{Luminosity function in $I$ for all measured stars 
in the NGC 3379 field.
In the uppermost graph, the number of stars per unit magnitude is plotted 
against $I$; the middle panel shows the numerically calculated 
first derivative $dn/dI$ of the LF, while the lower panel shows the numerical
estimate for the second derivative $d^2n/dI^2$. In all panels the data have
been convolved with a Gaussian smoothing kernel of $\sigma = 0.02$ mag.}
\label{lf} 
\end{figure}

We show the luminosity function in Figure \ref{lf}.  The version shown here has been 
smoothed with a Gaussian kernel of $\sigma_I = 0.02$ mag, although the result
is insensitive to the particular smoothing width.  Completeness
corrections are also quite unimportant here, since the $f=0.5$ completeness level
is considerably fainter than the well resolved top of the RGB.
In essence, we look for the maximum 
change in the LF slope near that point by using the 
numerically calculated first and second derivatives of
the LF (shown in the lower two panels of Fig.~\ref{lf} and referred to as
the ``edge response filter'' or ERF).
These show that the first sharp peak is at $I=26.10 \pm 0.10$, which we adopt
as the TRGB.

The distance modulus follows immediately once we apply a fiducial value
for $M_I(tip)$.  As we did in our NGC 3377 study, we adopt
$M_I(tip) = -4.05 \pm 0.12$ from the comprehensive photometric study
of $\omega$ Cen by \citet{bell04}, which is entirely consistent with
the range given by recent theoretical RGB models \citep[e.g.][]{sal02} 
depending on the details of the input stellar physics.
We therefore
obtain $(m-M)_I = 30.15 \pm 0.15$ for NGC 3379.  
This must be corrected for the foreground absorption 
of $A_I = 0.05\pm0.02$, giving a final TRGB distance measurement $\mu = 30.10 \pm 0.16$.
This result is entirely consistent with the previous TRGB 
measurements \citep{sakai97,gregg04} within their internal uncertainties.

Averaging our TRGB distance in with the SBF and PNLF measurements listed above,
and giving the three methods equal weights, we arrive
at an average $(m-M)_0 = 30.06 \pm 0.10$,
or $D = 10.2 \pm 0.5$ Mpc for NGC 3379.  
These three methods give a result which puts NGC 3379 near the average
for the Leo group as a whole, and is consistent with it being close to the
dynamical center of the group.  By comparison, our result for the
smaller elliptical NGC 3377 from exactly the same method 
placed it 0.10 mag more distant than NGC 3379.
This differential distance is just on
the margin of being significant relative to the internal uncertainty
of the TRGB method and suggests that the Leo group may have a minimum
line-of-sight ``depth'' of $\sim 1$ Mpc.

\section{The Metallicity Distribution}
\label{metallicity}

With the CMD and the distance calibration in hand, we are 
in a position to derive the 
metallicity distribution function  for the halo of this
galaxy.  To facilitate the most direct 
possible comparisons with other systems, 
we follow precisely the same method as in our previous studies
\citep{h02,rej05,har07}.  We place a 
finely spaced grid of RGB evolutionary tracks for 12-Gyr-old stars
\citep[the $\alpha-$enhanced tracks of][]{van00}
on the measured CMD, suitably registered to match the observed RGB
sequences for Milky Way globular clusters.  Interpolation within
the fiducial tracks is then carried out to estimate the 
heavy-element abundance $Z$ of each RGB star.
The details of this technique are described fully in \citet{h02} and we 
do not repeat them here. 
However, as before, we strongly emphasize that the metallicity scale is 
an {\sl observationally calibrated one} based on real globular clusters.
{\sl The theoretical models are used only to aid interpolation between
the observed sequences for real clusters.} 

We use the 12-Gyr models as a plausible age estimate for old halo stars
and globular clusters, while also realizing that for low-mass stars older than
$\sim 5$ Gyr the $(V-I)$ colors are only very weakly sensitive to
age \citep[e.g.][]{h99,rej05}.  If 
the stars in our target galaxy are actually younger
than 12 Gyr, then this method would slightly
underestimate their $Z-$abundances since the RGB locus shifts blueward
at lower age. But because the shift is only at the rate 
of $\Delta$log $Z \sim 0.1$ dex per 4-Gyr
age difference, the metallicity spread is by far the dominant effect 
in driving the large color
range that we see across the top of the RGB.

In Figure \ref{cmd_fiducial} we show the CMD with
the RGB tracks added.  
The two tracks shown as dashed lines at right are ones
for Solar ($Z=Z_{\odot}$) and $\simeq 3 Z_{\odot}$ metallicities;
both of these fall past the 50\% photometric completeness level in $V$
and thus imply that {\sl if} this remote outer part of the NGC 3379 halo
does contain any such stars, most would not be detectable in our data.
Considerably deeper exposures in $V$ will be needed to find them unambigously.

The derived MDF, plotted in conventional form as number of stars
per unit [m/H] = log$(Z/Z_{\odot})$,  is shown in Figure \ref{feh_3panel},
where we divide the sample into half-magnitude bins by approximate
luminosity $M_{bol}$.  
Fig.~\ref{feh_3panel} explicitly shows the MDF with, and without,
photometric completeness corrections. 
Any stars fainter than the $f=$50\% line in {\sl either} 
$F606W$ or $F814W$ have been
rejected from the sample, since at these levels the completeness
correction itself becomes dangerously large and the random and
systematic errors of the photometry increase rapidly.  
For all stars brighter than the 50\% cutoff, the completeness-corrected
samples (the open histograms in Fig.~\ref{feh_3panel}) have been constructed
by weighting each star individually as $(1/f)$ where $f = f_I \cdot f_V$
is the combined completeness fraction at its particular location in
the CMD.  For comparison, the unweighted MDF (based only on counting up
all stars with $f > 0.5$) is shown in the hatched regions.
The completeness corrections affect the shape of
the MDF histogram in an important way only for [m/H] $> -0.3$.  

The faintest of the three bins reaches into
the $I > 27$ magnitude range that is still likely to be affected to
some extent by field contamination (see the preceding discussion), but
any such contamination does not seem to have skewed the overall shape
of the MDF by comparison with the two brighter bins.  Nevertheless,
in the following discussion we use only the brightest ($M_{bol} < -2.5$)
part of the data, corresponding roughly to the uppermost magnitude
of the RGB.  

The shape of the MDF is a surprise.  The previous results from
other E galaxies including NGC 3377 \citep{har07}, NGC 5128 \citep{h02},
and also M32, a galaxy near the lower limit of the normal E sequence
\citep{gri96}, as well as the near-infrared NGC 3379 data of 
\citet{gregg04}, appeared to establish a pattern in which a large spread of
RGB metallicities is present but where the great majority of stars are
metal-rich with MDF peaks in the range $\langle$m/H$\rangle \simeq -0.7$
to $-0.3$ depending on galaxy luminosity.
However, both the distribution of the stars on the NGC 3379 CMD, 
and its transformed
version in Fig.\ref{feh_3panel}, are strikingly unlike any of the other systems.
The MDF is the broadest and flattest one
we have ever seen.  
Once the transformation from $(V-I)$ to [m/H] has been made, we find that 
this part of the  halo is
not dominated by either low-metallicity or high-metallicity components.
The mode of the distribution seems to be near [m/H] $\sim -0.5$, but
unlike all the other galaxies cited above,
there is really no interval in the MDF that is genuinely dominant.  
Neither can the MDF shape be described easily as ``bimodal'' as is the case
for almost all globular cluster systems in large galaxies, where roughly
equal numbers of clusters concentrate near [m/H] $\simeq -1.5$ and $\simeq -0.5$
\citep[e.g.][]{peng06,harris06}.
In addition, since the MDF is still not declining very rapidly at the
upper end where it hits the photometric completeness cutoff, it seems likely
that it actually continues up to and beyond Solar metallicity 
\citep{gregg04} and that
we are seeing only a lower limit to its full extent.  In the discussion
below, we will estimate more quantitatively how many more metal-rich 
stars we are likely to be missing from the complete MDF.

\section{Matching to Chemical Evolution Models}
\label{evolution}

To gain a bit more insight into the possible formation history
of NGC 3379, we next try to step
beyond the raw MDF into a chemical evolution model.

In our series of studies of NGC 5128  
we developed a simple, semi-analytic
chemical evolution model that has been applied successfully
to all the NGC 5128 fields, to NGC 3377 \citep{har07}, and to
the dwarf ellipticals \citep{h99,butler05}.  Very similar models have also 
been used for the halo of the Milky Way
\citep{prantzos03}, and the globular cluster systems of large
galaxies \citep{van04}, among other situations.  Briefly, in
this first-order model we envisage an
``accreting box'' in which a region
of initial gas mass $M_0$ turns itself into stars through a long
succession of star-forming episodes, during which more gas is
continuously flowing into the region.  Although in reality this
star-forming sequence will happen continuously, for numerical
calculation purposes we suppose it to happen in 
a series of small discrete timesteps $\delta t$.  
By hypothesis, the rate of gas infall is allowed to die away
with time, so that in the late stages of the
region's history, its chemical evolution asymptotically approaches
the classic ``closed-box'' or ``Simple'' model \citep{pagel75}.  
By carrying out a straightforward numerical integration, we
then compute the total number of stars at a given metallicity (that is,
the model MDF) once all the gas has been used up.
As we discuss in the papers cited above, this
model is an approximate description of what would be expected to happen
during hierarchical merging of a large set of initial, zero-metallicity 
gas clouds within which
star formation is taking place simultaneously as they merge to form
a bigger final galaxy.

In \citet{h02} we outline and justify the key assumptions in the model:
\begin{itemize}
\item{} The gas infall rate starts at a chosen
level and then dies away as an exponential decay with time. 
\item{} At each star formation step $\delta t$, the same 
fraction of ambient gas gets turned into stars (we adopt a
5\% conversion rate for purposes of the numerical calculations).
\item{} Each timestep assumes ``prompt mixing'', i.e. at each stage
the remaining gas in the region has a uniform $Z(gas)$.
\item{} The abundance $Z$ of the stars
forming at any given moment then equals the abundance 
of the gas
left behind by the previous steps, mixed with the 
new gas entering the region just before the next star formation
step occurs.
\item{} The ``effective yield'' $y_{eff}$ of the stellar nucleosynthesis
(the fraction of stellar mass that
is expelled back into the interstellar medium as enriched heavy
elements) is assumed to stay constant throughout the sequence.
\end{itemize}

The model has a minimum of three free parameters:  (1) the effective 
yield $y_{eff}$, which combines the effects of both the true 
nucleosynthetic yield $y$ and any SN-driven outflow that
drives gas out of the system \citep[cf.][]{binney98}; 
(2) the initial gas infall rate $(\dot M /M)_0$ relative
to the amount of gas initially present in the region; and (3) the
exponential decay time $\tau_2$ for the infall rate.  Other potentially
useful parameters include (4) an initial time period $\tau_1$
over which the infall rate $\dot M$ stays roughly constant; 
and (5) the heavy-element abundance $Z_{in}$ of the added gas.
The so-called closed-box or Simple Model is a special
case where we set $\dot M, \tau_1, \tau_2$
equal to zero, leaving only $y_{eff}$ as the single free parameter.

An extremely instructive way to study the match between model and
data is through the linear form of the MDF, which is the number of stars per
unit heavy-element abundance $Z/Z_{\odot}$.  In this graph, the
closed-box model would look simply like an exponential decline in
$dn/dZ$, the number of stars per unit heavy-element abundance.
In Figure \ref{zmodel1}, we replot the $Z-$histogram of our data along
with two particular cases of a closed-box model.
The models shown in Fig.~\ref{zmodel1} can match either the
high$-Z$ or low$-Z$ end of the data, but 
no single choice of $y_{eff}$ fits the entire run,
so the Simple model is not even approximately valid.
 
For NGC 5128, NGC 3377, and the dwarf ellipticals,
we found that although a closed-box evolution does not
fit them either, it is possible in each case to start from
primordial material $Z_{in} = 0$ and then to find an accreting-box solution
with reasonable choices of $y_{eff}$, $\tau_2$, and $\dot M_0$
that gives an excellent match to the data.
NGC 3379 is unlike all these previous cases.  Experimentation with
the accreting-box model shows that
{\sl no single chemical evolution sequence of this type}
can fit this MDF.

The next step is to try a multi-stage model.  We show one such solution
in Figure \ref{zmodel2} which provides reasonable success, and
which assumes that the formation process happened in two rather
distinct stages. The parameters for each stage are:
\begin{itemize}
\item{} {\sl Metal-poor component:} A closed-box model with $\dot M = 0$,
$y_{eff} = 0.1 Z_{\odot}$, and a truncation of the timestep sequence near
$Z = 0.2 Z_{\odot}$.  This truncation point is adopted as
the obvious point above which the simple model
with low yield $y_{eff} = 0.1 Z_{\odot}$ can no longer match the data
(see Figure 7).  We use a sharp truncation only for numerical simplicity;
a steep but smoother ramp-down at that point would work equally well.

\item{} {\sl Metal-rich component:} An accreting-box model with $Z_{in} = Z_0 
= 0.19 Z_{\odot}$ (that is, both the initial abundance and the infalling
gas have the same, nonzero enrichment), $y_{eff} = 0.5 Z_{\odot}$,
$\tau_1 = 6 \cdot \delta t$, and $\tau_2 = 7 \cdot \delta t$.
\end{itemize}

The example we show in Fig.~\ref{zmodel2} is meant only to be illustrative;
other sets of parameters can be found similar to these which also give 
plausible deconstructions of the MDF.  For an example of the accreting-box
model applied to a still more distinct two-stage model, see \citet{van04}
and their discussion of bimodal MDFs for globular cluster systems.
Several possible combinations of parameters are shown there, along
with a nonlinear statistical procedure for finding the best-fitting
parameters for an assumed model.  Their discussion shows, however, that
noticeably different model assumptions can lead to equally good combinations
of model parameters, and the only way to select among these is through
external physical constraints.  The only clear constraint we have for
the NGC 3379 data (within the context of the accreting-box models) is
the empirically well defined changeover between modes at $Z \simeq 0.2 Z_{\odot}$.
Given this, we find that the final adopted $y_{eff}$
is {\sl internally} uncertain by $\pm 10$ percent and the infall times
 $\tau_1, \tau_2$ by $\pm 2 \cdot \delta t$.
For the abundance of the infalling gas, $Z_{in}$ for the higher-metallicity
mode needs to be within
10 percent of $Z_0 = 0.2 Z_{\odot}$ to maintain the continuous transition
between the two modes.

If interpreted at face value, the model shown above suggests that NGC 3379   
underwent two fairly distinct epochs in its formation history.  
First was the buildup of a classic, metal-poor halo starting with 
pristine gas, a low
effective yield, and without much ``infall'',
roughly resembling what we
find in the Milky Way or in the dwarf ellipticals.  We speculate that the end of this
phase near $Z \simeq 0.2 Z_{\odot}$ may, perhaps, be connected with
the epoch of cosmological reionization that could have interrupted the first
rounds of star formation in the pregalactic dwarf population
around redshifts $z \sim 6 - 10$, including the metal-poor
globular clusters \citep[e.g.][]{santos03,rhode05}.  In this connection,
\citet{beasley02} have noted that a truncation redshift quite similar
to {\sl z(reionization)} is necessary to
produce a distinctly bimodal MDF for globular clusters in their
semianalytic model of galaxy formation.

The second major stage was the 
buildup of the metal-rich, bulge-like component (starting from gas that was
pre-enriched from the first phase?) and with a higher effective yield.
This second phase
continued long enough for the star-forming gas to enrich up to Solar abundance or higher.
The factor-of-five difference in $y_{eff}$ between the two stages
suggests that the ``halo'' star formation could have taken place in small potential
wells (pregalactic clouds or dwarf-sized satellites) where a high
fraction of the gas was lost to outflow; whereas the metal-richer
component could be made within a deeper potential well that could hold on
to much more of the gas.
For comparison, in \citet{h02} we found $y_{eff} \simeq 0.3 Z_{\odot}$ for 
the outer halo of NGC 5128 (a more massive giant than NGC 3379; see 
the Discussion section below),
while for the inner region of NGC 5128, 
we found $y_{eff} \simeq 0.85 Z_{\odot}$, which approaches the 
typical theoretically expected nucleosynthetic yield without gas loss.

We do not discuss here an alternate approach of building up the metal-poor
component of the halo completely by accretion of small, low-metallicity
satellites at later times.  Although there is clear evidence that these
kinds of events are happening in the Milky Way and other large galaxies,
evidence from the detailed abundance patterns of the heavy elements
\citep[which are different between the Milky Way halo stars and the
dwarf spheroidal satellites; see][]{venn04,font06,pritzl05} argues that the entirety of
the halo did not build by late accretion.

The model shown in Fig.~\ref{zmodel2} also gives us a way to estimate the
effects of photometric incompleteness on our measured MDF.  If we
extrapolate the same model past the observational cutoff $Z > 0.5 Z_{\odot}$
out to $Z \sim 2 Z_{\odot}$  \citep[the upper limit suggested by][]{gregg04}, 
we should add another $\simeq 13$\% to the entire population
of stars in the diagram. Said differently, we would be missing about one-quarter
of just the {\sl metal-rich} component alone because of our photometric cutoff.

A more model-independent way to check this estimate of the numbers of {\sl very }
metal-rich stars is to look for stars on our original images that are well
above our photometric limit in $I$, but below our cutoff in $V$.
From the {\sl allstar} file of measured objects in $F814W$,
1464 stars brighter than $I = 27$ were also measured in $F606W$,
survived the cuts for $\chi_{V,I}$ and stellarity, and thus
appeared in the final CMD.  But in addition to these, there are
$\simeq 260$ objects with $I \lesssim 27$ 
that were not matched successfully with anything in the $F606W$ image
and thus could be very metal-rich RGB stars.  These totals give
an {\sl upper limit} that our MDF could contain as 
many as 18\% more stars beyond our photometric cutoff,
entirely consistent with the model extrapolation described above.
In summary, we do not appear to be missing a major part of the MDF
at this location in the halo.

\section{Comparison with Globular Clusters}
\label{gcs}

The old-halo globular cluster (GC) population that is always present
in E galaxies gives a second way to assess the
metallicity distribution in the halo.  Recent wide-field photometric
studies of the GCs in NGC 3379 have been carried out by
\citet{whit03} and \citet{rhode04} which verify earlier results
that its GC population is quite small (unfortunately for our purposes).
\citet{rhode04} estimate that the total GC population
comprises only $N_{GC} \sim 270$ clusters, making the
specific frequency (number of GCs per unit galaxy
luminosity) $S_N = 1.2 \pm 0.3$.  This level is 3 to 4 times lower than 
the average for typical
large ellipticals in Virgo, Fornax, and other cluster environments,
but not unlike at least some other ellipticals in the ``field''
and small groups
\citep{har01}.  Despite their small numbers, the GCs display the normal
bimodal color and metallicity distribution that shows up almost
universally in large galaxies \citep[e.g.][]{har01,peng06,harris06}.
Using the $(B-R)$ color index for a total of 36 well measured
GCs, \citet{rhode04} deduce the presence of metal-poor and metal-rich subpopulations
that are fairly distinct from each other.  We have converted their
$(B-R)$ histogram into [Fe/H] with our own calibration based on 80
Milky Way globular clusters with low reddenings and measured
colors \citep{har96}, 
\begin{equation}{\rm [Fe/H]} \, = \, 3.13 (B-R)_0 - 5.04
\end{equation}
along with $E_{B-R} = 0.04$ for NGC 3379.  (Rhode \& Zepf do the
same to deduce the mean metallicities of each of the two modes,
but do not quote the actual conversion relation they used.)
We find that the blue GC mode is at [m/H] $\simeq$ [Fe/H] + 0.2 
$= -1.33$, and the red mode at [m/H] $\simeq -0.36$.
Both of these mean points are
internally uncertain to $\pm 0.2$ dex.  A double-Gaussian fit
to the histogram shows that $\simeq$79\% of the GC population is
in the blue mode and just 21\% in the red mode, consistent with
Rhode \& Zepf's estimates.  The internal uncertainties in both
the red and blue GC groups are high because of small-number
statistics, though the red side is
clearly the less certain of the two.

The natural question is to ask whether 
these two GC metallicity subgroups have any connection to the
two-mode field-star MDF we discuss in the previous section. In Figure \ref{fehgcs},
we compare the two types of objects directly.  For [m/H] $\lesssim -1$, the
GC and RGB distributions match up well, consistent with the idea that
the metal-poor clusters and field stars formed at the same time.
If so, the continued formation of both may have been truncated at
the same time (see discussion above).
For the more metal-rich half of the MDF, the numbers of GCs are very
much smaller and it is not yet clear whether their underlying distribution
by metallicity has the same shape.  It is notable, however, that the 
proportions of the two metallicity subgroups are very
different, with the field halo including many more metal-rich
objects relative to the metal-poor ones.  
This fundamental observed difference between the field stars and
halo clusters is a type of ``specific frequency problem'' 
that appears to be quite a general issue in giant galaxies,
and has not yet found a compelling explanation
\citep[see][for further discussion]{har01,h02,bea03}.  The lowest-metallicity
massive star clusters in some way formed at very high efficiency
relative to the field stars.

A recent study of the GCs by \cite{pie06} uses Gemini/GMOS spectra 
to measure ages, metallicities and abundance ratios for a sample of about
two dozen GCs over the full range of metallicities.  Although the sample
size is small, they find that the clusters are uniformly old ($> 10$ Gyr),
as is the case in the Milky Way.

The possibility that NGC 3379 is actually not a true elliptical, but
a nearly face-on S0 galaxy, has been raised on the basis of the
details of its light distribution \citep[e.g.][]{cap91,ss99}.
Interestingly, the globular clusters provide some circumstantial
evidence consistent with such an interpretation:  the very low
specific frequency $S_N \simeq 1.2$ would not be unusual for an S0 or a large
disk galaxy, but is certainly on the extreme low end for true
ellipticals.  Another relevant piece of evidence is discussed by
\citet{pie06} and is based on kinematics.  The planetary nebulae in
the galaxy show a velocity dispersion that gradually declines with 
galactocentric distance \citep{rom03}, whereas the velocity dispersion
in the globular cluster system \citep{pie06} stays roughly constant with $R_{gc}$, as it would
in a normal dark-matter halo with isotropic orbits.  One way to reconcile
the PN velocities with the GCs would be to suggest that the PNe have
progressively increasing radial anisotropy outward.  However, if the
galaxy is an S0 that we see nearly face-on, the PNe might be more associated
with a disklike population and thus have a lower dispersion along our
line of sight, by contrast with the more spherically distributed GCs
\citep[see][for additional discussion]{pie06}.

\section{The Metallicity Gradient}
\label{gradient}

It is notable that the inner halo fields studied with {\sl NICMOS}
by \citet{gregg04} showed no significant numbers of low-metallicity stars,
whereas our outer-halo field shows a large number of them.  This
comparison, and the chemical evolution argument made in the previous
section, suggests that we should look more closely for 
traces of a {\sl metallicity gradient}.  Fortunately, the $200''$ width
of the ACS/WFC field is
large enough to span a radial range 27.7 kpc to 38.1 kpc from
the center of NGC 3379.  Across this one field, do we see any changes in
the relative numbers of metal-poor and metal-rich stars?

The answer is yes.  In Figure \ref{xy2}, we show the positions of
the brighter ($26 < I < 27.3$)
measured stars on the image, where we have subdivided them into
the same two major groups that were identified from the entire
MDF:  a ``blue'' metal-poor population with $Z < 0.2 Z_{\odot}$ 
([m/H] $< -0.7$),
and a ``red'' group with $Z > 0.2 Z_{\odot}$.  The blue group includes the
obvious dE-like RGB that defines the low-metallicity half of
the chemical evolution model in Fig.~\ref{zmodel2}.  Recall that the
upper part of the image (large $y-$values) is the east side of the
frame, closest to NGC 3379.  For the red group, a very obvious
density gradient appears, while the blue group is more evenly
spread.  

We show these two subpopulations again in Figure \ref{profile},
plotted as the number density $\sigma$ of stars per unit area as a function of
position $y$ or, alternately, projected radius $R_{gc}$ from the center of
the galaxy.  Approximate power-law profiles can be matched to each
one, but it is the difference between the two that is striking. The blue RGB 
is well described by $\sigma \sim R^{(-1.2 \pm 0.7)}$, whereas the red RGB
population needs
a much steeper gradient near $\sigma \sim R^{(-6.0 \pm 0.6)}$ to match 
the data.\footnote{To determine these radial curves we select stars in
the range $I < 27.3$.  It should be noted that this is not the same limit
as used above for the MDF, and the total numbers of stars in each group
should thus not be compared.  We use the latter magnitude cut only to
estimate the radial slopes.}
For both groups combined, the overall gradient is $\sigma \sim R^{-4.5 \pm 0.5}$.

The large-scale surface brightness (SB) distribution of the galaxy, with the ACS
field position marked on it, is shown in Figure \ref{devauc}.  The standard
$r^{1/4}$ profile determined by \citet{devauc79} and extended by 
\citet{cap90} is $\mu_B = 14.076 + 3.0083 R^{1/4}$ for $R$ in arcseconds
and $\mu_B$ in mag/arcsec$^2$; note, however, that for $\mu_B > 28$ the
measurements are quite uncertain \citep[see][]{cap90}.
At our field position of $r \simeq 650''$, we expect 
$\mu_B \simeq 29.3$.  Replotted as surface brightness versus
log $R$, a straight line in $\mu$ versus $R^{1/4}$ takes on a steadily 
steeper logarithmic slope
for increasing $R$, and at our field position, the profile would predict that
the surface brightness should vary approximately as $I \sim R^{-6}$, consistent
with the power-law exponent that we observe for the red RGB
stars.  

A third way of representing the metallicity gradient is shown in
Figure \ref{cmd2}.  Here, we plot CMDs separately for the upper half
($y > 2000$ px, eastern side) and the lower half ($y < 2000$ px, western 
side).  The west side, which is the one further from the center of NGC 3379, has far fewer
metal-rich giants, and its CMD begins to resemble rather closely that
of a typical metal-poor dwarf elliptical \citep{han97, butler05}.

Our data indicate that at radii $R \lesssim 500''$ the starcounts (and
thus the halo surface brightness) should be completely dominated by the
higher-metallicity stars. These are what we see in the 
published SB profile.  However, if the SB measurements could be extended
accurately outward beyond $R > 800''$, 
we should see a flattening of the
profile to something closer to $I \sim R^{-2}$ or even less as the metal-poor
component takes over.  Surface brightness measurements at such low
levels are exceptionally difficult, however, and the most effective
way to continue them outward is likely to be precisely the method we have
used here: that is, through starcounts of the resolved RGB population.

The presence of a metallicity gradient also allows us to reconcile our
data rather easily with those of \citet{gregg04}:  if the metal-rich
component rises inward as steeply as we see it do, then their 
innner-halo CMDs should have been dominated by high-metallicity
stars, as in fact they are.  The small {\sl NICMOS} field and thus
the smaller absolute numbers of stars tends to exaggerate their 
prominence as well.  If we use the power laws from Fig.\ref{profile}
to scale the expected number densities of the bluer and redder RGB
components from the inner edge of our ACS field at $R = 9.3'$ 
inward to
(e.g.) the middle NICMOS field at $R = 4.5'$, then we would predict
that the redder RGB stars should outnumber the bluer ones at the
same limiting $I$ luminosity by at least
10 to 1 there, making them very hard to detect.  In fact, such a
scaling is likely to be extremely uncertain since it does not account
for the flattening off of the surface brightness profile of either
component (see again Fig.~\ref{devauc}).
\citet{gregg04} suggest that they find a small metal-poor component
amidst the dominant metal-rich one, but this conclusion rests heavily
on an extremely accurate understanding of their photometric uncertainties
in the faintest 1 mag of their data (see their Fig.~12).
The best way to determine the true radial profile of the intriguing
metal-poor component will be by further deep observations at
selected and widely spaced radii.

Lastly, in Figure \ref{fehsplit} we show a direct comparison
between the two MDFs in the eastern and western halves of
our field.  In the upper panel of the Figure, the two are
superimposed on each other, but normalized
to the same {\sl total} number of stars more metal-poor
than a rather arbitrarily chosen dividing line
at [m/H] $ = -1$.  When normalized
this way, the eastern half has an excess of metal-rich stars,
shown in the lower panel which plots the difference between
the two halves.  The excess metal-rich component
must have a mean metallicity of at least $\langle$m/H$\rangle 
> -0.5$, typical of the bulges or disks of large galaxies.
We can set only a lower limit to it, since a significant fraction
of it is likely to extend above our photometrically imposed
cutoff near [m/H](max) $\simeq -0.2$.

Our choice of field location turns out to be more fortunate than
we initially anticipated.  It is, apparently, {\sl just} at the radial
range that we study in which the metal-rich component is rapidly
dying out, and in which the metal-poor halo is starting to emerge
clearly from underneath it.  We have been able to see both components at once.
If we had selected a field pointing $\sim 3' - 5'$ either farther
out or farther in, our CMD would have been dominated by only one
of these components and the nature of the population gradient would
not have been as clear.  The shallow
surface density dependence $\sigma \sim R^{-1.2 \pm 0.7}$
of the metal-poor halo component is, if anything, even flatter than that expected for
an NFW potential well, which should converge to a projected density
$\sigma \sim R^{-2}$ in its outer regions \citep{nfw97}, although
the difference is within the uncertainty of measurement.  The slope
would be at least roughly
consistent with the idea (see above) that the low-metallicity
stars formed in pregalactic clouds that were widely dispersed
within the dark-matter halo before the main epoch of hierarchical
merging and dissipative collapse.

A last bit of evidence worth noting is that the metal-rich
{\sl globular clusters} (discussed in the previous section)
have a different spatial distribution than the metal-poor clusters.
The data from \citet{rhode04} (see their Fig.~10) show that
no red GCs are found beyond $R_{gc} \simeq 12'$, just at the 
radial distance of our ACS field where we see the metal-rich
RGB stars rapidly declining.  By contrast, the blue GCs continue
on outward detectably to $R_{gc} > 20'$ (60 kpc).  This comparison
is consistent with the often-expressed view \citep[e.g.][]{har01,bea03,h02}
that the metal-rich GCs formed with the bulk of the gE that is
predominantly metal-rich.  

Our data from NGC 3379, however, add new
evidence to support the complementary view that 
the {\sl metal-poor stars and GCs} both belonged 
to an earlier stage at which the
protogalactic gas clouds were more widely distributed.
What remains unexplained about this epoch is the relatively
large proportion of GCs to metal-poor halo stars.

If in fact the galaxy is actually a nearly face-on S0 rather than
an E1 (see above), our interpretation would have to be modified
to the view that the moderately metal-rich {\sl disk} component
dies out steeply near our field location, again leaving the 
metal-poor halo to continue outward. 

\section{Discussion}
\label{discussion}

In other large E galaxies for which MDFs have been derived based on
{\sl direct photometry of halo stars}, no radial metallicity 
gradients in the MDFs have been seen.  What distinguishes NGC 3379 from these
others?  

The most obvious possibility to raise is that, in these other
systems, the existing photometry has not reached far enough out
into the halo {\sl relative to the effective radius of the spheroid}.  In 
NGC 3379, the 
transition from metal-rich dominance to metal-poor dominance
occurs at the $R_{gc}$ range from $10 R_e$ to $13 R_e$, where the effective radius
is $R_e = 0\farcm93 = 2.8$ kpc.  For NGC 5128, the four fields
studied so far range from $1.4 R_e$ to $7.3 R_e$ ($R_e = 5.5$ kpc);
and for NGC 3377, the field surveyed covers $1.5 R_e$ to $5.3 R_e$
($R_e = 3.3$ kpc).  
Thus in NGC 3379 we have sampled the halo about twice as far out
{\sl in units of} $R_e$ as in these other systems.

\begin{figure} 
\figurenum{5} 
%\plotone{f5.eps} 
\plotone{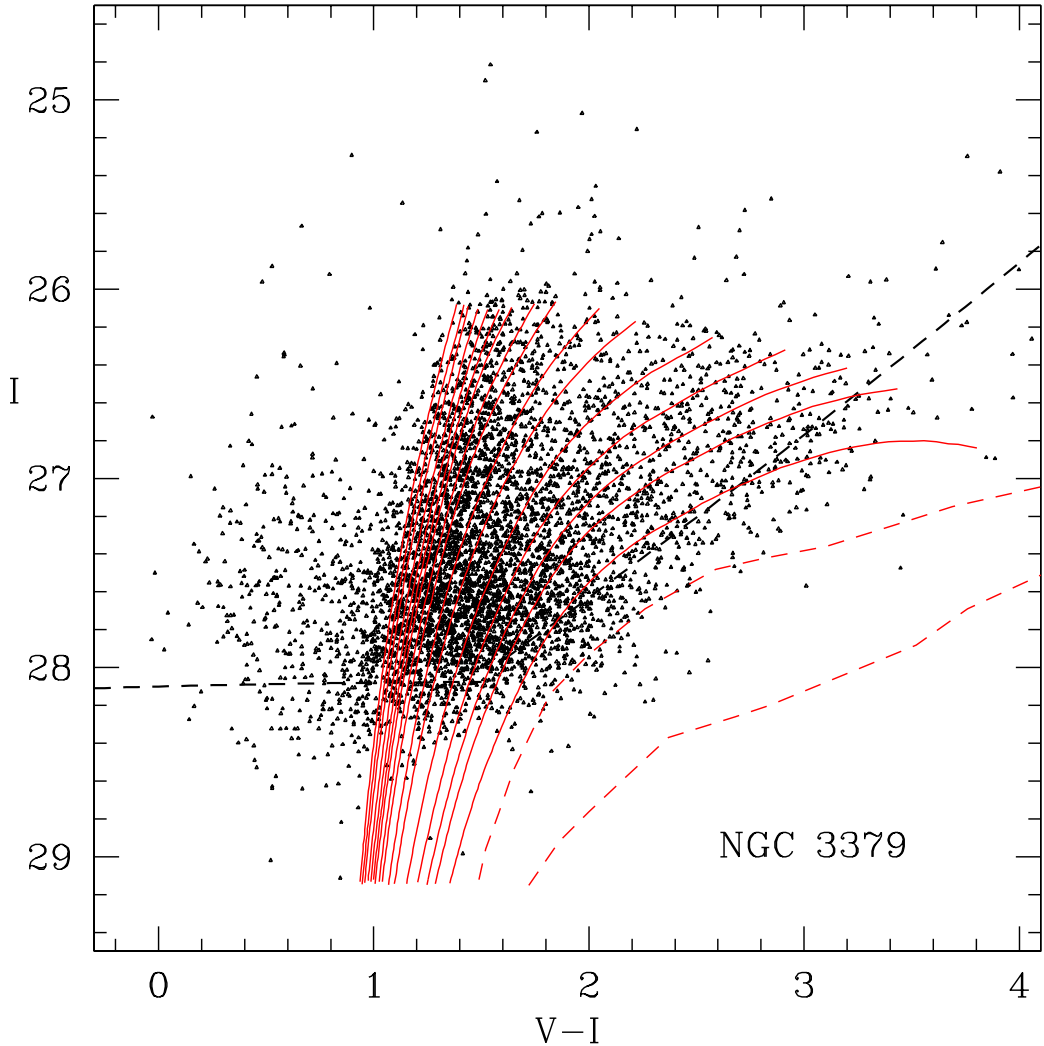} 
\caption{Color-magnitude diagram for the NGC 3379 red giants, with model
red-giant tracks superimposed on the data points.  All tracks are for
ages of 12 Gyr, but differ in metallicity roughly in steps of 
$\Delta$[Fe/H] $\simeq 0.1$.  These are the same set of tracks from
VandenBerg et al.~(2000) we have used for our previous studies of the
NGC 5128 and NGC 3377 halos, supplemented by two 
metal-rich tracks generated from old Milky Way star clusters
(see Harris \& Harris 2002).  The total metallicity range for
the entire grid extends from log $(Z/Z_{\odot}) = -2.0$
to $+0.4$.  The placement of the
tracks assumes a foreground reddening $E(V-I) = 0.02$ and a
distance modulus $(m-M)_I = 30.15$ as derived in this study.
The 50\% photometric completeness limit is shown as the pair
of dashed lines.}
\label{cmd_fiducial} 
\end{figure}

\begin{figure} 
\figurenum{6}
\plotone{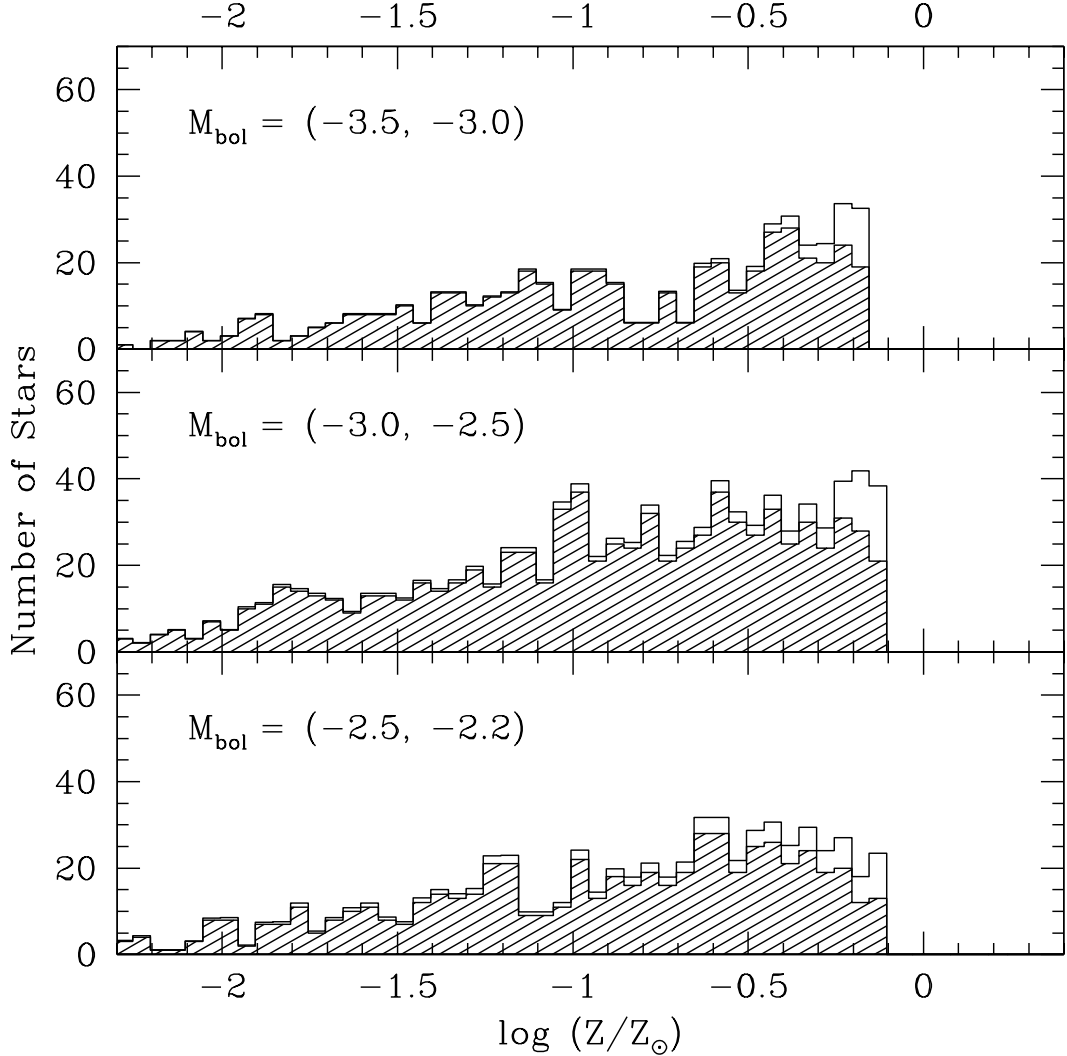} 
\caption{Metallicity distribution function for the halo red giants, divided
into three approximate luminosity bins.  This division tests for any systematic
errors in the interpolation routine or the placement of the evolutionary tracks.
The {\sl shaded regions} show the MDF uncorrected for photometric completeness,
while the higher {\sl unshaded regions} show the full completeness-corrected MDF.
Stars fainter than the 50\% photometric completeness lines
(see previous Figure) are not included in the sample, which sets
the abrupt high-metallicity cutoff to the observed MDF.}
\label{feh_3panel} 
\end{figure}

\begin{figure} 
\figurenum{7} 
\plotone{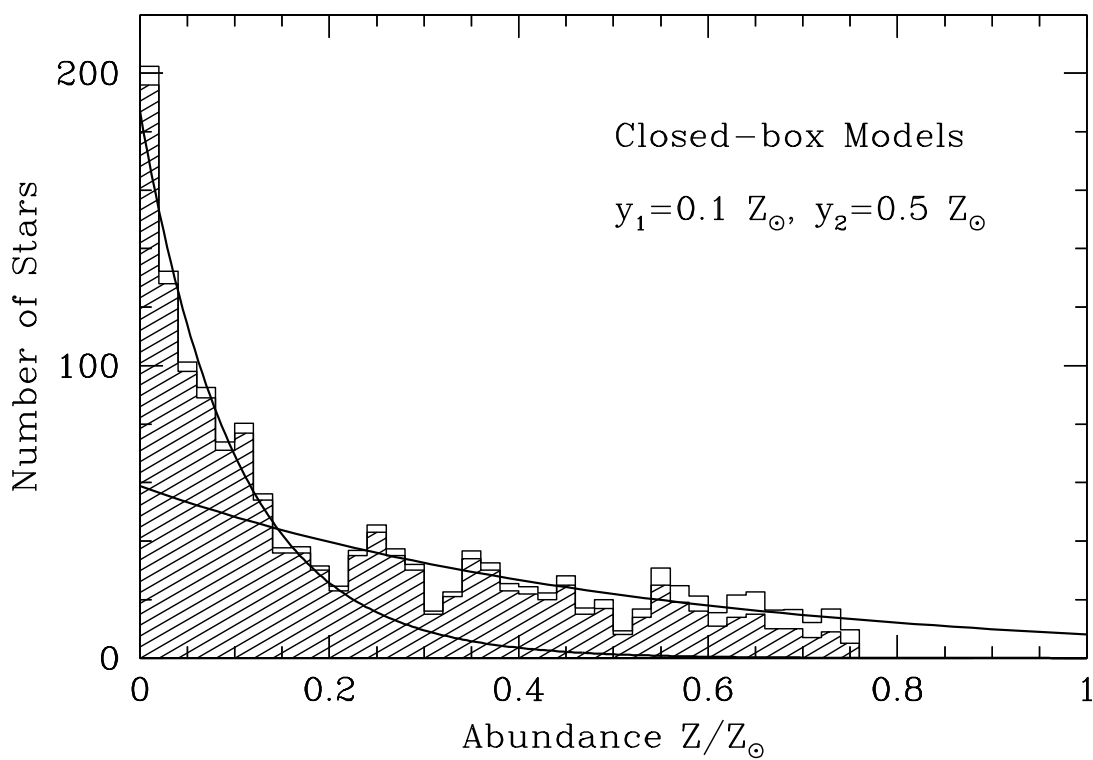} 
\caption{Histogram of heavy-element abundance distribution, plotted in
linear form as number of stars per unit $Z/Z_{\odot}$.  Stars from only the
brightest two bins in the previous diagram ($M_{bol} < -2.5$) are used here.
The predicted $N(Z)$ distributions for two separate closed-box models are
shown as the solid lines.  The steeper one has 
effective yield $y_{eff} = 0.1 Z_{\odot}$, and the shallower one has
$y_{eff} = 0.5 Z_{\odot}$. Each one is a simple exponential decay curve
with $e-$folding scale given by the effective yield.
}
\label{zmodel1} 
\end{figure}

\begin{figure} 
\figurenum{8} 
\plotone{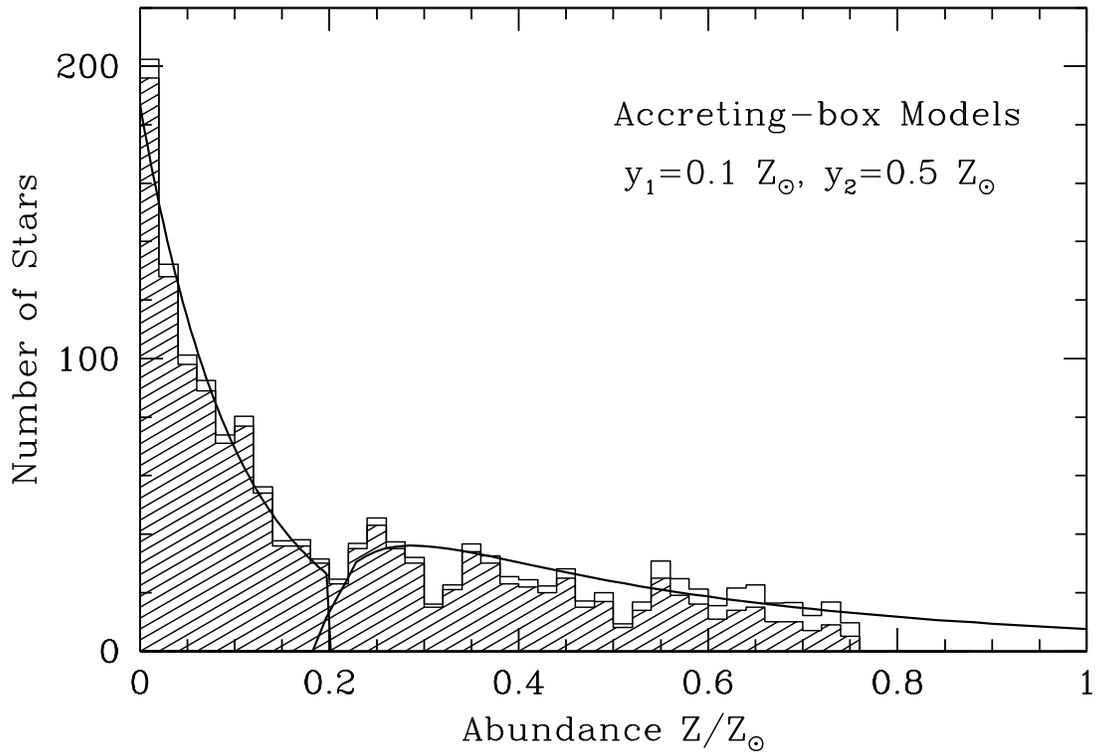} 
\caption{The same abundance distribution as in the previous figure,
but now matched with a two-stage chemical evolution model as described
in the text.  The same effective yields are used as in the previous
figure, but with different assumptions about the gas infall rate
and abundance.  The low-metallicity component ($Z < 0.2 Z_{\odot}$) has
an effective yield $y_{eff} = 0.1 Z_{\odot}$, while the
high-metallicity component ($Z > 0.2 Z_{\odot}$) starts with pre-enriched
gas and has an effective yield 5 times higher.}
\label{zmodel2} 
\end{figure}

\begin{figure} 
\figurenum{9} 
\plotone{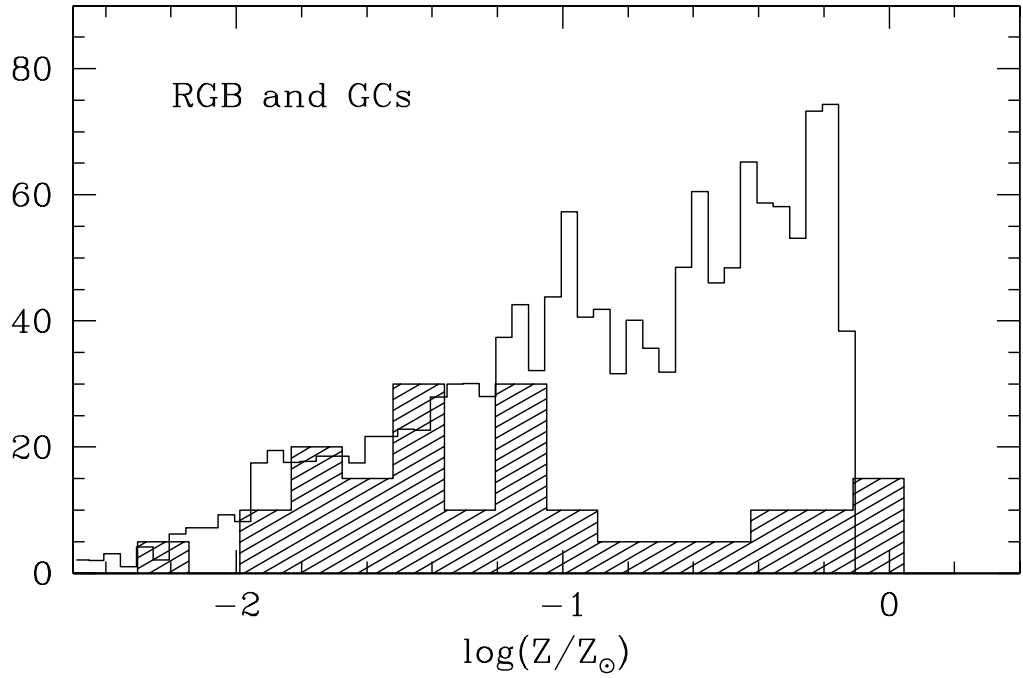} 
\caption{Metallicity distribution for the bright RGB stars, including
the same data as in the previous graph, now compared with the metallicities
of the globular clusters in NGC 3379 (data from Rhode \& Zepf 2004).
The RGB stars are shown as the open histogram and the GCs as the shaded
histogram.  For clarity we have multiplied the numbers of GCs per bin
by a factor of 4 to normalize the metal-poor numbers to the RGB stars.
}
\label{fehgcs} 
\end{figure}

\begin{figure} 
\figurenum{10} 
\plotone{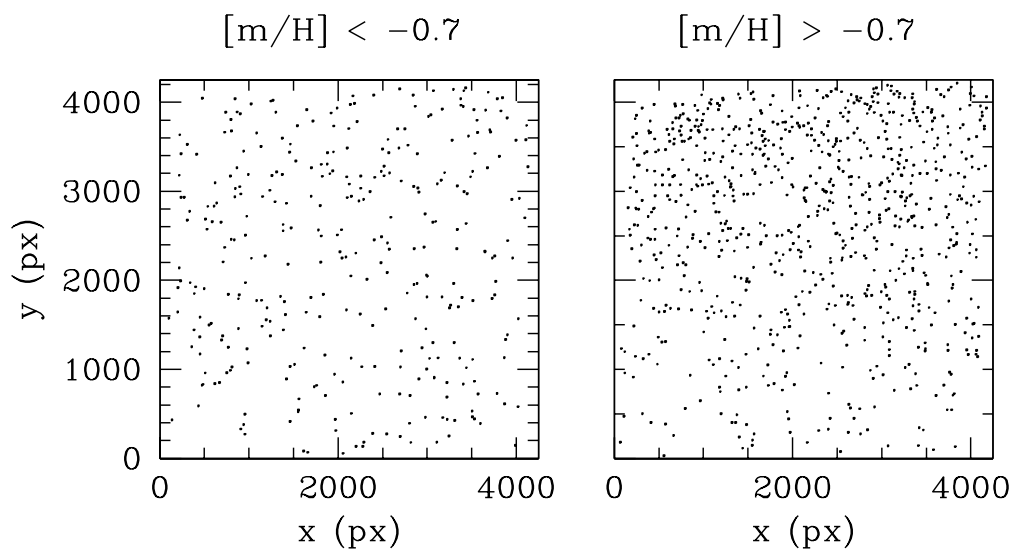} 
\caption{Positions of the bright stars on the ACS image, in the 
magnitude range $26.0 < I < 27.3$.  The center of NGC 3379 is
off the diagram at the top.  The left panel shows the
blue, metal-poor RGB stars ([m/H] $< -0.7$, equivalent
to $Z < 0.2 Z_{\odot}$) while the right
panel shows the red, metal-richer giants ([m/H] $> -0.7$).  The
red population exhibits a much stronger gradient in number 
density across the frame.
}
\label{xy2} 
\end{figure}

\begin{figure} 
\figurenum{11} 
\plotone{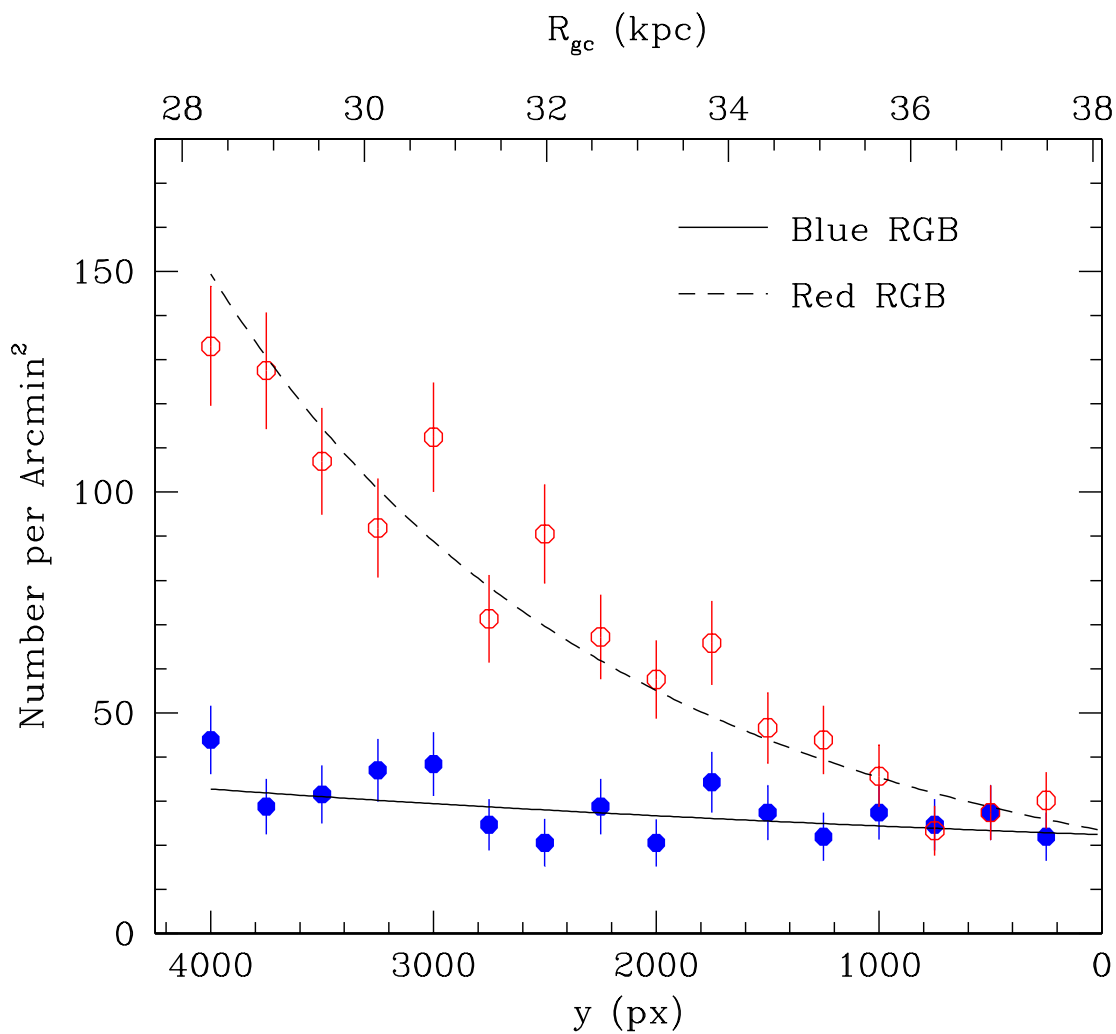} 
\caption{Number density $\sigma$ (number of stars per arcmin$^2$)
as a function of position on the image.  The solid symbols and
solid line show the data for the blue, metal-poor stars 
with [m/H] $< -0.7$ plotted
in the previous figure.  The open symbols and dashed line show
the same data for the red, metal-rich group with [m/H] $ > -0.7$.
The numbers along the top border of the plot give the projected
distance $R_{gc}$ (in kpc) from the center of NGC 3379.  The blue population
density falls off with radius as $\sigma \sim R^{-1.2}$ (solid
curve) while the red population falls off as $\sigma \sim R^{-6.0}$
(dashed curve).
}
\label{profile} 
\end{figure}

\begin{figure}
\figurenum{12}
\plotone{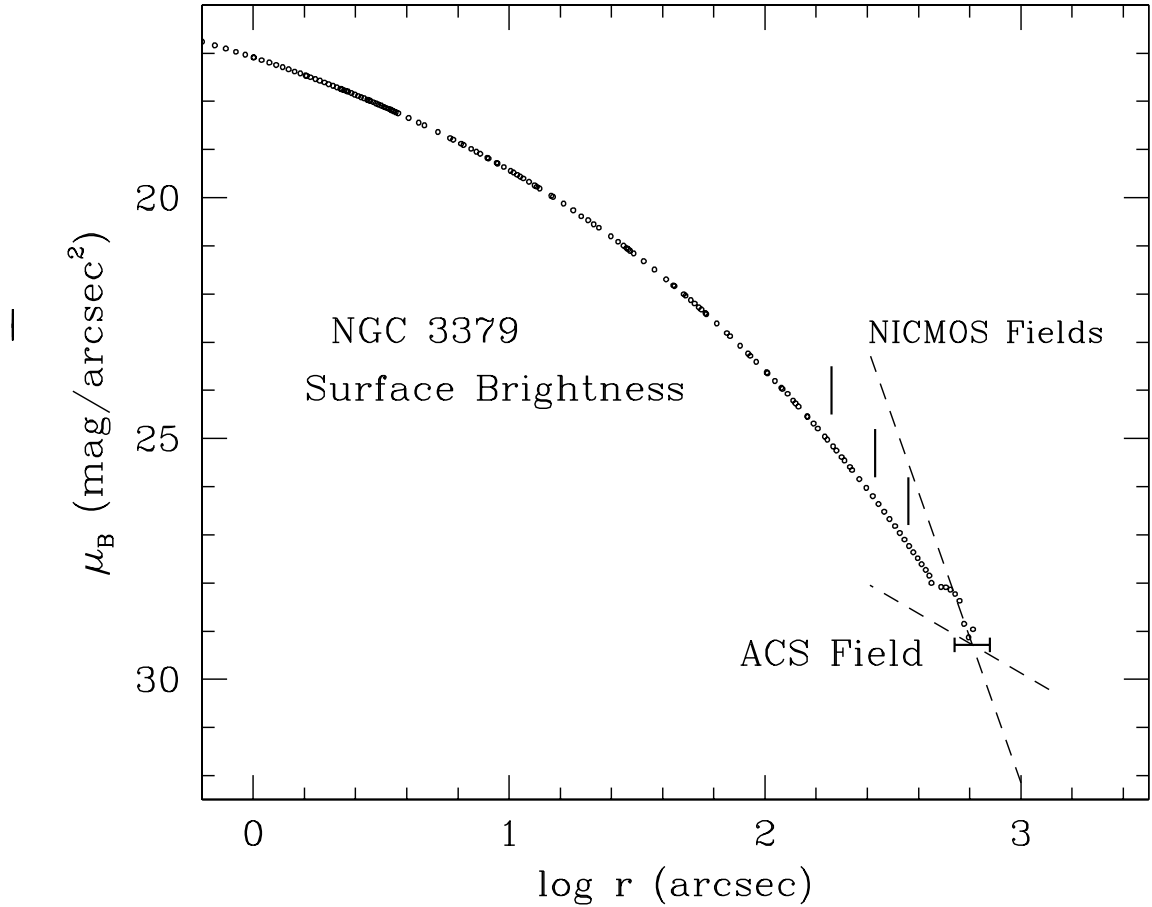}
\caption{Surface brightness in $B$ along the EW axis for NGC 3377,
with data from \protect\citet{devauc79} and \protect\citet{cap90}.
The $B-$band surface brightness in magnitudes per square arcsecond
is plotted against projected radius in arcseconds.  The surface
brightness measurements
are shown as the dots, while the location of our ACS field is shown
by the small horizontal bar at lower right (the length of the bar
shows the radial extent of the field).  The three vertical tick marks
show the locations of the NICMOS fields studied by Gregg et al. (2004).
The steeper dashed line
drawn through the ACS field has a power-law exponent of $-6.0$, showing
the falloff rate of the redder, more metal-rich RGB stars in the previous diagram.
The shallower dashed line has an exponent of $-1.2$, representing
the bluer, metal-poor RGB stars.
}
\label{devauc}
\end{figure}

\begin{figure}
\figurenum{13}
\plotone{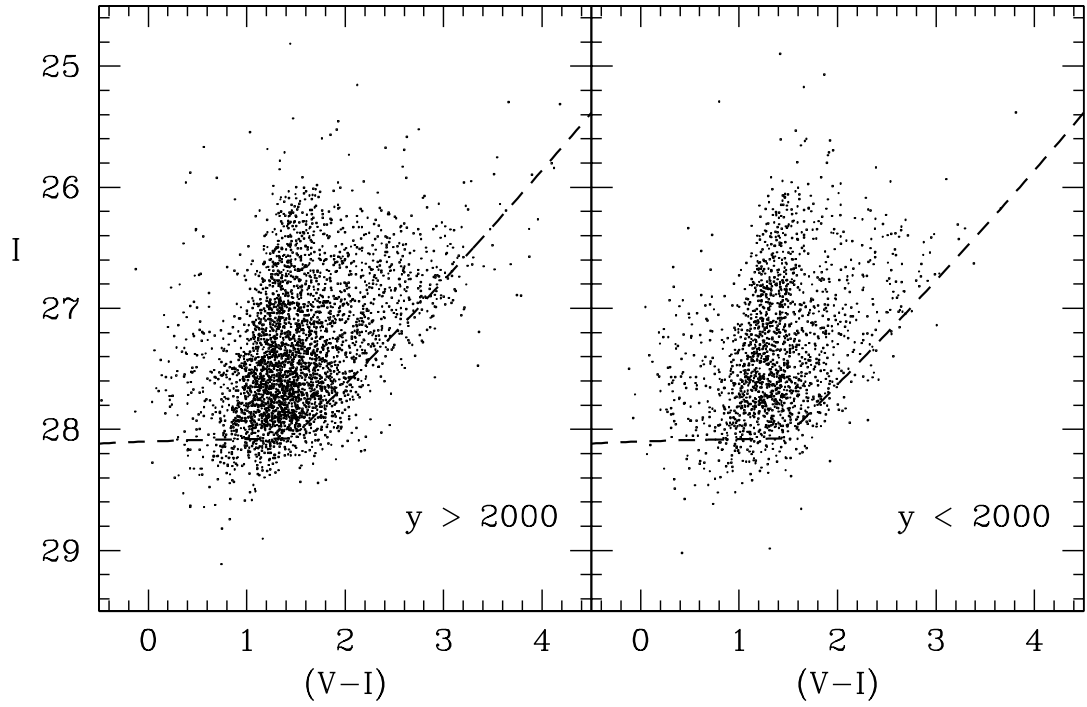}
\caption{Color-magnitude diagrams for the upper (eastern) half
of the frame ($y > 2000$ px) and the lower (western) half
($y < 2000$ px).  The eastern half is the one closer to
NGC 3379 and it has the higher proportion of red, metal-rich stars.
}
\label{cmd2}
\end{figure}

\begin{figure}
\figurenum{14}
\plotone{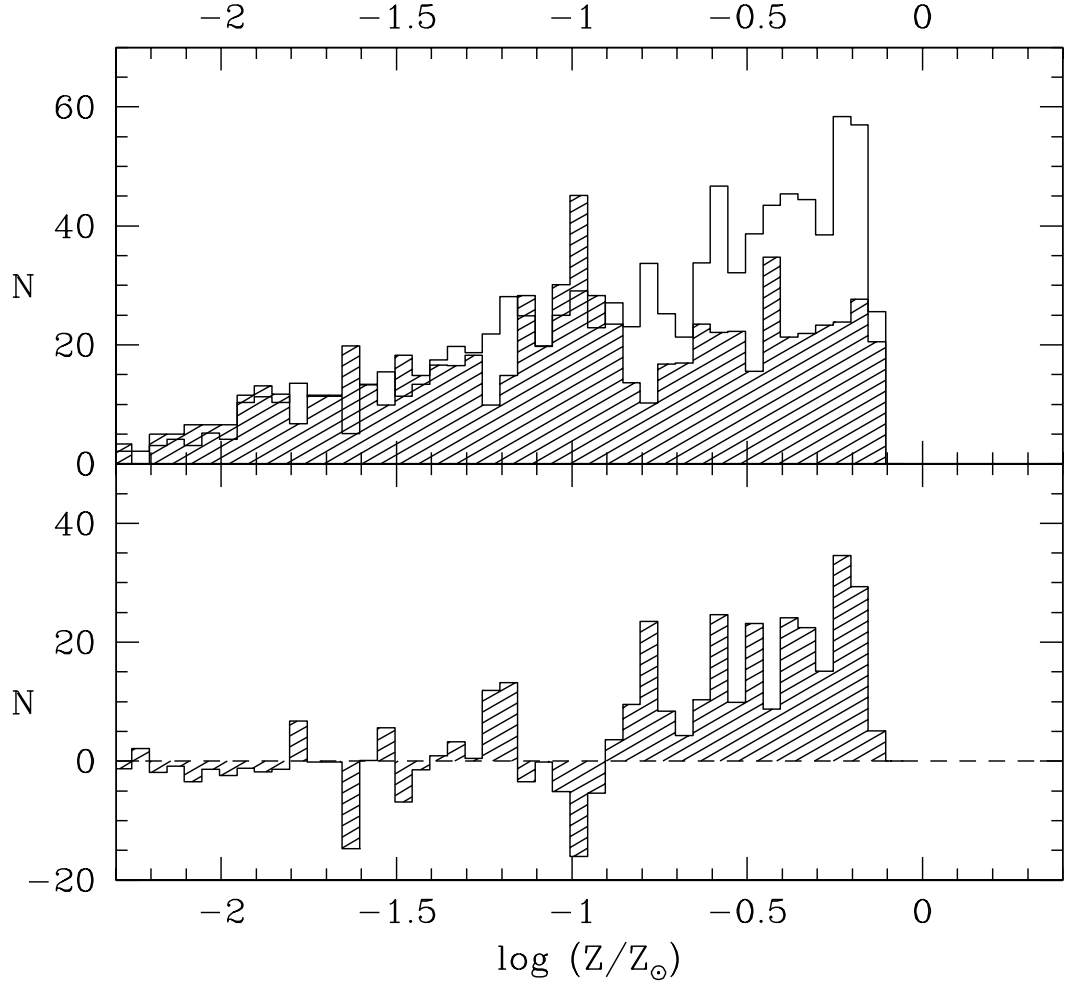}
\caption{{\sl Top panel:}  Metallicity distribution for the 
stars in the upper half of the image ($y > 2000$ px, shown as
the open histogram) and the lower half ($y < 2000$ px, shown
as the shaded histogram.  Both graphs have been normalized
to the same total number of stars with [m/H] $ < -1$.
{\sl Bottom panel:}  Residual plot of the difference between
the two histograms in the top panel.  The upper half of the
image has an excess of metal-rich stars ([m/H] $> -0.8$)
relative to the metal-poor ones.
}
\label{fehsplit}
\end{figure}

The most directly comparable observations in the literature may instead
come from those of a giant {\sl disk} galaxy, M31.  Several recent
studies have begun to trace the stellar density distribution and
metallicity distribution of the M31 stars outward as far as
$R = 165$ kpc in projected radius 
\citep[][and references cited there]{irw03,guh06,kal06}.
For $R \lesssim 30$ kpc, the MDF is moderately metal-rich, like those
in the large ellipticals we have discussed above, and with no
obvious metallicity gradient.  Beyond $R \sim 60$ kpc, however, the
MDF becomes much more metal-poor with a mean near [Fe/H] $\simeq -1.3$ to $-1.5$
that resembles the Milky Way halo \citep{kal06}.  These authors state
that the ``transition zone'' between a bulge-like metal-rich population
and the metal-poor halo starts near 30 kpc.  Since $R_e = 4.7$ kpc for
M31 ($R_e = 20\farcm8$ from RC3), 
their conclusion is essentially that the ``pure halo'' component
becomes dominant only for $R \gtrsim 10 R_e$, entirely consistent with what
we have found.
In addition, the fact that
no metallicity gradient shows up for either $R < 30$ kpc or $R > 60$ kpc
indicates that the two components are each fairly homogeneous, and that
the overall metallicity gradient is the result of changing proportions
of the two with radius.  We invite the reader to compare Kalirai et al.'s
color-magnitude diagram (their Figure 6) 
with our Figure \ref{cmd_fiducial}:  even though there are far
fewer stars in their M31 diagram, its overall pattern  
relative to the grid of RGB isochrones strongly resembles 
what we find for NGC 3379.

A more well known galaxy with an extended metal-poor halo is of
course the Milky Way.  The spatial distribution for metal-poor stars and
globular clusters follows a three-dimensional profile $\phi \sim r^{-a}$
where $a \simeq 3.0 - 3.5$ \citep{har01,kin94,slu98,mai05} and
continues outward well past $r > 50$ kpc, similar to what we are beginning
to find specifically for M31 and NGC 3379.   
In a few other nearby disk galaxies that are edge-on, evidence for   
low-metallicity, very low-density halos is also now beginning to
emerge \citep[e.g.][]{set07,tik06} with the same method of resolved
stellar photometry.  These studies are finding that the halo
component becomes noticeable at heights of $\simeq 5$ kpc or more
above the disk plane.
If these metal-poor haloes are common features of galaxies, they
mostly likely have remained undetected until recent years
simply because of their extremely low
surface brightnesses at large galactic radii.  For example, in 
the Milky Way, the metal-poor halo has $\mu$ = 30 mag arcsec$^{-2}$
at the Solar radius \citep{binney98}, a level of faintness that is
extremely challenging even for the most meticulous surface-brightness
photometry. The alternate route (at least for nearby galaxies)
of resolved-star photometry can reach much fainter equivalent levels.

Yet another, though more indirect, window into the 
outer-halo stellar populations of galaxies may be found in the
intracluster stellar light found in clusters of galaxies.  
These free-floating stars, unbound to any particular member galaxy, are stripped
from infalling galaxies during tidal interactions.  \citet{mur04} 
find that not only do intracluster (IC) stars {\sl not} represent a
random sampling from cluster galaxies, but that they tend to be older
than bound stars, a result supported by \citet{som05}.
\citet{wil04} also find an age disparity
between their bound and unbound stellar populations, and in their
simulations of galaxy clusters, they trace the origin of stars that
end up in the intracluster population.  They find that galaxies of
all sizes contribute stars to the ambient IC population and that stars
are preferentially pulled from the outer and lower-metallicity
parts of their stellar distributions.  Although a study of Virgo IC
stars by \citet{dur02} finds stars of predominantly
intermediate metallicity, there is evidence of substantial structure
in the IC light in Virgo \citep{mih05}
and its unbound stellar populations are most likely not yet well-mixed.

\citet{wil07} obtained ACS data on an intracluster field in
Virgo, and find an MDF roughly similar to ours.  A comparison of our
Fig.\ref{cmd_fiducial} with the CMD in their Figure 7 
reveals that the metallicity distribution
of these Virgo cluster stars is rather similarly spread over the full
abundance range. \citet{wil07} argue that a noticeable proportion of
younger stars is present in this sample in addition to a basic old
population with a broad MDF.  While some mixing is expected in the IC
population, the fact that this intracluster MDF resembles those in
the outer halos of NGC 3379 and M31 might itself be indirect evidence
that large galaxies in general possess metal-poor halos, because the
tidally stripped material will come preferentially from those outer
parts of large and intermediate galaxies.

\begin{figure}
\figurenum{15}
\plotone{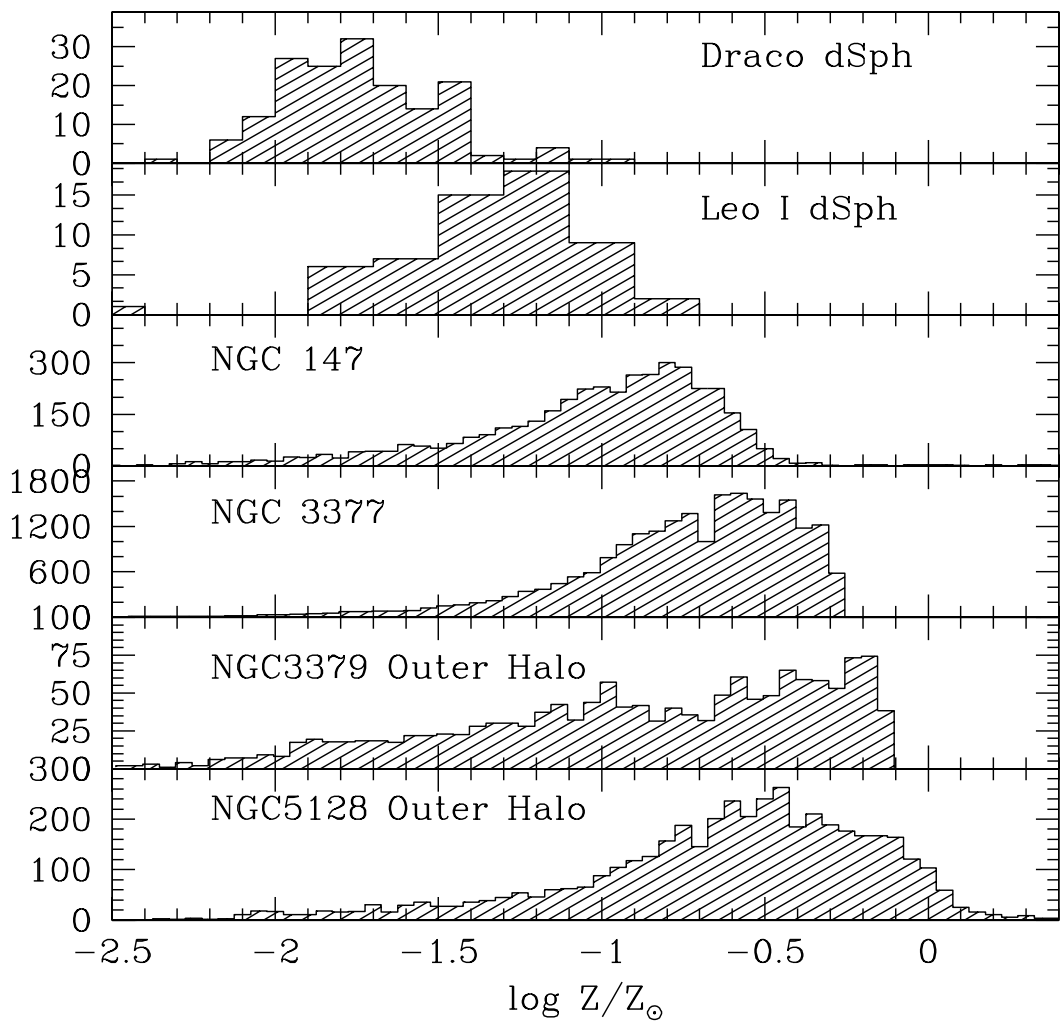}
\caption{Metallicity distribution functions for 6 elliptical 
galaxies, including the Local Group dwarf spheroidals Draco and Leo I,
the dwarf elliptical NGC 147, the Leo
intermediate-luminosity elliptical NGC 3377, and the giants
NGC 3379 and NGC 5128.  For NGC 3377 the field shown is centered
at galactocentric radius $\sim 4 R_e$; for NGC 3379, $\sim 12 R_e$;
and for NGC 5128, $\sim 7 R_e$.  See text for discussion.
}
\label{6galaxy}
\end{figure}

A comparison of halo-star
MDFs in E galaxies, from dwarfs to giants, can now be made and
is shown more explicitly in Figure \ref{6galaxy}.  These include:
the Local Group dwarf spheroidals Draco ($M_V^T = -8.8$; Mateo 1998) 
and Leo I ($M_V^T = -11.9$), with spectroscopic and photometric metallicity data
from \citet{far07} and \citet{koch07};
the Local Group dwarf NGC 147 ($M_V^T = -15.6$), with
photometric $(V,I)$ RGB data from \citet{han97}; the intermediate-luminosity
Leo member NGC 3377 \citep{har07}; our data for NGC 3379; and the outermost NGC 5128
halo fields at $R_{gc} = 40$ kpc \citep{rej05}.  
We use Draco and Leo I only as representative cases of dwarf spheroidals
that allow us to exhibit as large as possible a range of host galaxy sizes.
In the latter four cases, exactly the
same interpolation code and RGB model grid has been used to derive
the MDF.  For the two dwarf spheroidals, with very low luminosities and
distinctly lower mean metallicities than the others, the MDFs are
as presented in the papers cited above.
The galaxies are shown in order of increasing total luminosity.
Note, in particular, that NGC 5128 is a giant twice as massive
as NGC 3379; dynamical mass measurements to comparably large radii give
$M \simeq 4 \times 10^{11} M_{\odot}$ for NGC 3379
\citep{ber06} and $M\simeq 10 \times 10^{11} M_{\odot}$ for NGC 5128
\citep{kaw07}.

The sequence in Fig.~\ref{6galaxy}
clearly shows up as a progressively increasing mean
heavy-element abundance (keeping in mind that both the NGC 3377 and NGC 3379
data are truncated 
by photometric incompleteness for $Z > 0.6 Z_{\odot}$ and that the true
numbers of stars at Solar metallicity and beyond are not accurately known).
By contrast, the other four
galaxies are more deeply sampled and
less affected by incompleteness.   
A secondary feature worth noting is the shape of the MDF at very
low $Z$ in NGC 147 and 3377, in which the number of stars
per unit $Z$ {\sl rises} steeply for $Z < 0.1 Z_{\odot}$ before turning
over and declining.  This MDF feature at low $Z$ can easily be 
interpreted as a signature of infall
of primordial, relatively unenriched gas
during the earliest stages of star formation.  By contrast, the low$-Z$ end
of the NGC 3379 MDF has relatively much larger numbers of stars,
closely matching a closed-box model with no need to invoke infall (see discussion
above).  This MDF would be consistent with rapid
very early formation, even before the major stages of hierarchical
merging began.  But it is not yet clear how much of this difference in MDF shape
between NGC 3379 and the other galaxies could be due to the
rather different locations in the halo that each one has been studied,
and whether or not the metal-rich and metal-poor subcomponents that we
propose actually do belong to different evolutionary stages.

The sequence of host galaxies shown
in Fig.~\ref{6galaxy} allows us to exhibit almost the entire observable
range of E galaxy types, with potential-well 
masses from $\sim 10^7 M_{\odot}$ up to
$10^{12} M_{\odot}$.  We miss only the ``supergiant'' or cD-type galaxies
with still higher masses.  
Nevertheless, the sequence we can now put together from existing data
already covers a large dynamic range.
In our simple chemical evolution model, 
the location of the peak or mode of the MDF in the sequence shown
is determined by the effective yield $y_{eff}$.  The observable trend
for $y_{eff}$ to 
increase systematically with the depth of the host potential well
of the galaxy as a whole, is highly suggestive of the ability to hold onto
the gas in the region through many successive rounds of star formation,
which is expected to be larger in the more massive galaxies.

It remains to ask what the origin of the {\sl low}-metallicity, diffuse halo
component is likely to be.  Buildup of the outer halo by
accretion of metal-poor satellite dwarfs is a much-discussed 
option, especially since traces of such events are being found for
the Milky Way and M31 \citep[][among many others]{maj93,maj96,new02,iba05,guh06}.
Part of the evidence for such accretion events involves 
the presence of tidal streams, clumps,
and overall patchiness of the light distribution.  The smoothness and
regularity of the halo profile around NGC 3379 may therefore argue
against accretion, unless most of these events happened so long
ago that all such features have had time to diffuse completely.
However, in that case the semantic distinction between halo buildup by
accretion, versus an ordinary sequence of early hierarchical merging,
becomes somewhat blurred \citep[e.g.][]{bek01}.

The formation of large E galaxies by major mergers of disk galaxies
is also a well established interpretation.  In such mergers, the majority
of the halo stars in the merged product are the moderately metal-rich
stars in the former disks of the progenitor galaxies, which can be launched
out to large radii \citep[e.g.][]{bekki03} and leave an extended
halo with little 
overall metallicity gradient.  However, these merger models also indicate
that at {\sl very} large radii (40 kpc and beyond), the metal-rich component
should die away and the ``true'' metal-poor halo stars of the progenitor
disk galaxies should become relatively more noticeable \citep[see][]{bekki03}.

The possible role of formation by the more conventional route of dissipational
collapse, starting from a widely spread distribution of primordial
gas clouds, should not be ignored.  Existing dissipative-collapse 
models \citep[e.g.][for a recent representative example applied 
to E galaxies]{ang03}
are, however, pointed at recovering the metallicity gradients 
that are observed in the brightest inner halos and bulges of 
E galaxies \citep[e.g.][]{dav93},
which should arise during the main stage of the bulge and
spheroid formation. 
The remote, metal-poor regions could have an earlier
origin.  A broad-based simulation for disk-galaxy
formation which includes the
large-scale halo \citep{bek01} suggests that its outskirts were
formed by a combination of dissipative collapse and accretion
of satellites, leaving a present-day halo with a roughly spherical
distribution and a projected radial profile slope $a \simeq 2.5$.
As noted earlier, the expected slope at large radius might be $a \simeq 2$,
the profile of the dark-matter potential well \citep{nfw97}.

Much more observational work can obviously be done to test the
ideas discussed above, both in the nearest
systems such as NGC 5128 and the Leo galaxies, and in other systems. 
Our prediction is essentially that large galaxies in general will reveal 
extensive, metal-poor halos 
that become the dominant component at scale radii $R \gtrsim 12 R_e$.
Additional tests of the stellar number densities and MDFs at
different radii in these same galaxies would be of great interest.
Another set of target galaxies that would in many ways be optimum for
this type of deep, metallicity-sensitive stellar photometry would
be the Virgo cluster galaxies at $d = 16$ Mpc.  Virgo holds the largest
collection of E galaxies of all types in the local universe, and at
just one magnitude more distant than the Leo group, their red-giant halo
stars will be within reach of the HST ACS and WFC3 cameras without
prohibitively long exposure times \citep{wil07}.  In the search for
the extended metal-poor halos in giant galaxies that might represent
their earliest formation stages, however, attention should also be
paid to relatively more isolated systems where there will be less
concern about confusion from intracluster stellar light generated by
later dynamical evolution.  It is clear that with the right tools,
this observational field is just beginning.

\section{Summary}

We have used deep HST/ACS images of an outer-halo
field in the giant  
elliptical galaxy NGC 3379, a central member of the Leo group, to derive the 
metallicity distribution function of its old red-giant stars.  
A total of 5300 stars were
measured over a radial region extending from $9\farcm2$ to $12\farcm7$ 
(equivalent to 28 to 38 kpc, or 9.9 to 13.6 $R_e$) 
projected distance from the galaxy center.  
The $(V,I)$ data reach deep enough to reveal almost 2 magnitudes
of the red-giant branch and show what appears to be a purely old
population with little evidence for younger stars.

The MDF for this region of the halo
shows two major features that make it distinctly unlike
other E/S0 galaxies including
the Local Group dwarf ellipticals,
the intermediate-luminosity Leo member NGC 3377, and the Centaurus 
giant NGC 5128: first, 
the MDF for the NGC 3379 halo is extremely broad and flat, with many stars 
at every interval in [m/H] and only a gradual rise towards higher
metallicity.  Second, a {\sl metallicity gradient} is clearly
detectable across our ACS field.   
The higher-metallicity stars
([m/H] $> -0.7$) fall off steeply as $\sigma \sim R^{-6.0}$, while the
lower-metallicity component shows a much shallower profile 
$\sigma \sim R^{-1.2}$.  We suggest that our 
target field is just at the transition point
where, in essence, the high-metallicity component of the galaxy
reaches its edge and the shallower lowest-metallicity component becomes
the dominant one by default.

In order to successfully match this
extremely broad MDF, we find that a distinct two-stage accreting-box
chemical evolution model is necessary, whereby the metal-poor,
diffuse halo belonged to a rapid initial formation phase strongly
resembling a simple closed-box model with very low effective
yield $y_{eff} \simeq 0.1 Z_{\odot}$ and truncated at abundance
$Z \simeq 0.2 Z_{\odot}$.  By contrast, the metal-rich component
(the centrally concentrated spheroid or, possibly, the nearly
face-on disk if NGC 3379 is actually an S0)
requires an accreting-box type of model and an effective
yield 5 times higher.

Our target field is at $R = 12 R_e$, twice as far out 
in units of effective radius as in any
of the other galaxies that we have surveyed with this technique
of deep stellar photometry.  We believe that this is the key factor allowing
us to see the transition point between the metal-rich spheroid and
the more extensive and sparser metal-poor halo.
Some support
for this view comes from the recent observations of the outer halo
of M31, in which a metal-poor halo also emerges past $\sim 10 R_e$.
If NGC 3379 is indeed a representative case, we predict that 
large E/S0 galaxies {\sl in general} will have diffuse, 
very low-metallicity halo components, but
that photometry at radii $R \simeq 10 - 15 R_e$ will be necessary
to find them.

Finally, we summarize a series of arguments pointing to the possibility
that NGC 3379 may be an S0 galaxy rather than an E1.  These arguments
involve the internal dynamics and isophotal contours, the globular cluster
specific frequency, and the kinematic difference between the globular
clusters and the planetary nebulae.  None of these bits of evidence are
individually strong, however, and the possibility remains ambiguous.

\acknowledgements

WEH, GLHH, and EMHW thank the Natural Sciences and Engineering Research Council of 
Canada for financial support.  
ACL acknowledges Support for this work through grant 
number HST-GO-09811.01-A from the Space Telescope Science Institute.
We are indebted to James Taylor and Hugh Couchman for instructive 
conversations.

\clearpage

\begin{deluxetable}{ccccccccc}
\tablecaption{Basic Parameters for NGC 3379 \label{basics}}
\tablewidth{0pt}
\tablehead{
\colhead{Parameter} &  \colhead{Value} \\
}

\startdata
$Type$ & E1 \\
$\alpha$ (J2000) & $10^h 47^m 49\fs6$ \\
$\delta$ (J2000) & $12\fdg 34' 54''$ \\
$v_r$ (helio) & 911 km s$^{-1}$ \\
$A_V$ & 0.08 \\
$(m-M)_0$ & $30.06 \pm 0.10$ \\
$V_T^0$ & 9.20 \\
$M_V^T$ & $-20.85$ \\

\enddata

\end{deluxetable}

\clearpage

% 
% Figures 
% 

\clearpage

\clearpage

\clearpage

\clearpage

\clearpage

\clearpage

\clearpage

\clearpage

\clearpage

\clearpage

\clearpage

\clearpage

\clearpage

\clearpage

\end{document}